\documentclass{vgtc}                          

\graphicspath{{figures/}{pictures/}{images/}{./}} 
\usepackage[toc,page]{appendix}

\usepackage{times}                     

\usepackage{tabu}                      
\usepackage{booktabs}                  
\usepackage{lipsum}                    
\usepackage{mwe}    

\usepackage{mathptmx}                  

\vgtcinsertpkg
\def\checkmark{\tikz\fill[scale=0.4](0,.35) -- (.25,0) -- (1,.7) -- (.25,.15) -- cycle;}

\usepackage{tikz}
\newcommand*\circled[1]{\tikz[baseline=(char.base)]{
            \node[shape=circle,draw,inner sep=1pt] (char) {#1};}}

\title{See or Recall: A Sanity Check for the Role of Vision in Solving Visualization Question Answer Tasks with Multimodal LLMs}

\author{Zhimin Li\thanks{e-mail: zhimin.li@vanderbilt.edu}\\ %
        \scriptsize Vanderbilt University %
\and Haichao Miao\thanks{e-mail: miao1@llnl.gov}\\ %
     \scriptsize \centering Lawrence Livermore National Laboratory %
\and Xinyuan yan\thanks{e-mail: xyan@cs.utah.edu}\\ %
     \scriptsize University of Utah %
\and Valerio Pascucci\thanks{e-mail: pascucci@sci.utah.edu}\\ %
     \scriptsize University of Utah %
\and Matthew Berger\thanks{e-mail: matthew.berger@vanderbilt.edu}\\ %
     \scriptsize Vanderbilt University %
\and Shusen Liu\thanks{e-mail: martha.stewart@marthastewart.com}\\ %
     \parbox{1.4in}{\scriptsize \centering Lawrence Livermore National Laboratory}}

\abstract{
Recent developments in multimodal large language models (MLLM) have equipped language models to reason about vision and language jointly.
This permits MLLMs to both perceive and answer questions about data visualization across a variety of designs and tasks.
Applying MLLMs to a broad range of visualization tasks requires us to properly evaluate their capabilities, and the most common way to conduct evaluation is through measuring a model's visualization reasoning capability, analogous to how we would evaluate human understanding of visualizations (e.g., visualization literacy).
However, we found that in the context of visualization question answering (VisQA), how an MLLM perceives and reasons about visualizations can be fundamentally different from how humans approach the same problem.
During the evaluation, even without visualization, the model could correctly answer a substantial portion of the visualization test questions, regardless of whether any selection options were provided.
We hypothesize that the vast amount of knowledge encoded in the language model permits factual recall that supersedes the need to seek information from the visual signal.
It raises concerns that the current VisQA evaluation may not fully capture the models' visualization reasoning capabilities.
To address this, we propose a comprehensive sanity check framework that integrates a rule-based decision tree and a sanity check table to disentangle the effects of "seeing" (visual processing) and "recall" (reliance on prior knowledge).
This validates VisQA datasets for evaluation, highlighting where models are truly "seeing", positively or negatively affected by the factual recall, or relying on inductive biases for question answering.
Our study underscores the need for careful consideration in designing future visualization understanding studies when utilizing MLLMs.
} 


\begin{document}



\maketitle
\section{Introduction}
Recent advancements in multimodal large language models (MLLM) have facilitated their ability to perceive and understand images, charts, and diagrams \cite{yang2023dawn}, effectively allowing them to \textit{see}.
The capability of processing multimodal inputs has been explored in many general-purpose vision-related benchmarks \cite{li2024seed, xu2023chartbench, wu2024}. 
This new visual understanding capability of language models has also inspired exciting research directions in the visualization community, such as investigating an LLM's visualization reasoning abilities \cite{bendeck2024empirical, choe2024enhancing,li2024visualization} or developing autonomous visualization agents \cite{liu2024ava}.
Specifically, the visualization understanding capability has been increasingly studied \cite{bendeck2024empirical, li2024visualization,hong2025llms} in the context of visualization literacy \cite{lee2016vlat, pandey2023mini, nobre2024reading}.
These evaluations typically assume that MLLMs behave like humans, requiring them to first read and understand the given visualization, and then answer relevant questions based on their interpretation of the visual information provided in the prompt.

\begin{figure}[t]
\centering   
    \includegraphics[width=0.9\linewidth]{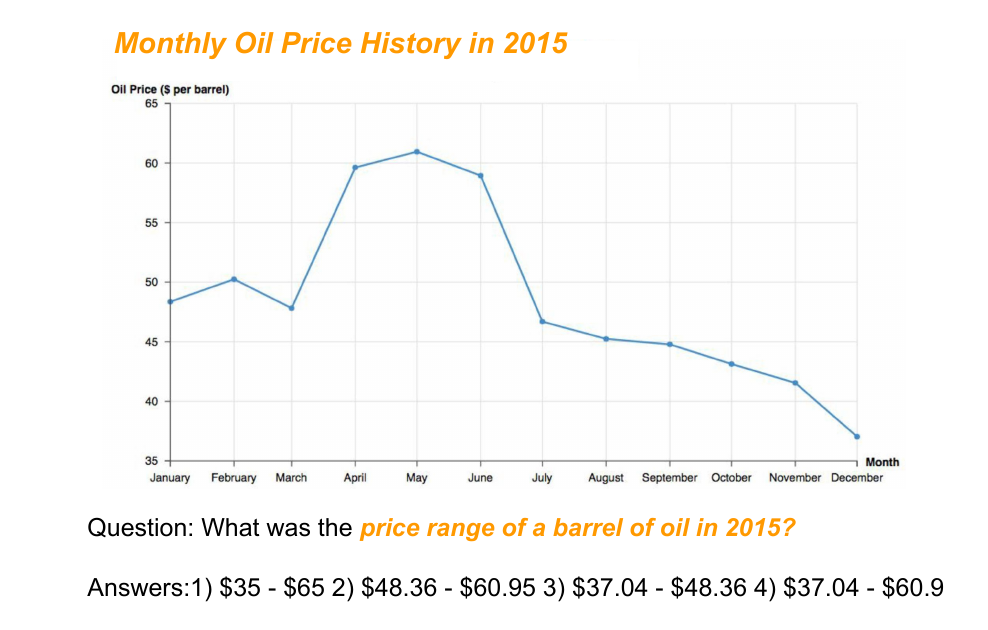}
    \caption{An example from the VLAT dataset where the highlighted text "oil price in 2015" provides context for MLLM that can trigger factual recall, allowing the model to answer correctly even without visual input.}
    \vspace{-4mm}
\label{fig:example}
\end{figure}

However, this assumption does not consider the possibility that MLLMs might not need to interpret a visualization to answer questions correctly.
Instead, they might rely on factual \textit{recall} from their training data in deciding on an answer.
This implies the model might discard the visual encoding of data, and rather attend to the question/answer text, if not the rendered text of the visualization in the form of titles, labels, and axes.
For example, Figure~\ref{fig:example} presents a question from the popular Visualization Literacy Assessment Test (VLAT) dataset designed to evaluate human visualization literacy.
When the same question is posed to the GPT-4 model without providing the accompanying visualization, the model still answers correctly and provides an explanation:
\textit{``The correct answer is:
4) \$37.04 - \$60.95.    In 2015, oil prices fluctuated between approximately \$37 and \$61 per barrel due to various market conditions, including oversupply and reduced demand.''}
Such an explanation implies that the model contains prior knowledge on economic data, and thus, the line plot might not be necessary for the model to give a correct answer.

This behavior raises critical questions: Are MLLMs actually ``Seeing'' the visualization to answer the question or simply ``Recalling'' the knowledge from training data to complete the task?
What role does visual input play in visualization question-answering (VisQA) tasks when evaluating MLLMs?
Instead of assuming that MLLMs process visual information like humans, our study aims to reveal the different factors influencing the MLLM's decision-making processes and to understand the role of vision in these tests. We propose a guideline for more robust and rigorous evaluation practices.
To understand these behaviors, we investigate the MLLM model's responses to the different VisQA datasets.
We observed that the MLLM can correctly answer a significant portion of questions without ever seeing the corresponding visualizations, indicating that the surprising example of Figure \ref{fig:example} is not the exception.
This observation raises the concern of what we are measuring in VisQA evaluation.
Do we have a correct interpretation of the evaluation results?
How do we know whether there is a chance the model may simply obtain the answer through factual recall instead of correctly interpreting the visualization as the test intends?

To gain a better understanding of these questions, we design a comprehensive framework of sanity checks for disentangling the impact of ``Recalling'' and ``Seeing'' for MLLM evaluation.
The framework starts with a decision-tree-based diagram on a single question to cover all outcomes of different ablation scenarios and categorize them into failure/success cases. The result of the decision tree is summarized into three cases of failing the sanity check.
The design is further extended to a probability sanity check table for general validation of a dataset.  
The evaluation of our methodology goes through each sanity check failing case with  VLAT~\cite{lee2016vlat} dataset, and then studies how our framework reveals the problem of counterfactual no-factual VisQA in the VLATforge dataset. We strengthen our with much larger datasets VILA~\cite{cui2024promises} and ChartQA~\cite{masry2022chartqa}.
Our experiment demonstrates that ``Recalling'' can potentially lead to overestimating or underestimating a model's performance on visual tasks. Meanwhile, this problem cannot be easily fixed through prompt engineering.

Our study presents a cautionary tale for researchers examining the visualization understanding capabilities of MLLMs and provides general guidelines for improving the robustness of such evaluations.
The findings also highlight a broader issue in MLLM's vision evaluation and raise fundamental questions about how to assess their vision capabilities properly. 
Overall, we summarize our main contributions as follows:





\begin{itemize}

    \item We demonstrate the widespread critical issue wherein MLLMs can answer visualization reasoning questions correctly without actually interpreting the visualizations, instead relying on factual recall from their extensive training data in the visualization reasoning context with multiple datasets (\textit{Sections 3}).
    
    \item We design a comprehensive and novel sanity check framework that utilizes a decision-tree-base abstraction to distill three validated cases and a sanity check table to verify the effects of ``recalling'' and ``seeing'' in MLLMs' decision-making processes (\textit{Section 4}).

    \item We experiment with the proposed sanity check framework that provides a robust methodology for evaluating MLLMs’ visualization reasoning ability over four diverse datasets covering factual context, non-factual context, multiple choice selection, and concise response Q\&A (\textit{section 5}).

    
    
\end{itemize}

 
    

\section{Related Work}





\textbf{Visualization competency test.}
Human comprehension of visualizations can vary widely based on one's education and experience in reading visualizations.
Börner et al.~\cite{borner2016investigating} assessed 273 museum visitors of varying ages and educational backgrounds, many of which struggled to interpret these visualizations, suggesting that information visualization is not widely accessible.
Efforts to improve visualization literacy have led to various competency tests.
Alper et al.~\cite{alper2017visualization} examined how visualization is taught in elementary schools and developed a web application to enhance children's visualization skills.
E. Firat et al.~\cite{firat2020treemap} focused on treemap literacy to evaluate human performance with treemap visualizations.
The VLAT~\cite{lee2016vlat} is a well-known test for measuring visualization literacy, comprising 12 visualizations and 53 questions.
Pandey and Ottley~\cite{pandey2023mini} condensed this into a Mini-VLAT with 12 questions.
Nobre et al.~\cite{nobre2024reading} replicated the VLAT study to provide deeper insights into the challenges of visualization interpretation.
As VLAT consists of a variety of visualization designs and tasks, scoring high on this test is an indicator of good visualization comprehension for a human.
Yet we should not assume a MLLM reasons about visualizations in the same way as a human, and this in part motivates our diagnosis of VisQA for MLLMs.

\begin{figure*}[t]
\centering  
    \includegraphics[width=0.32\linewidth]{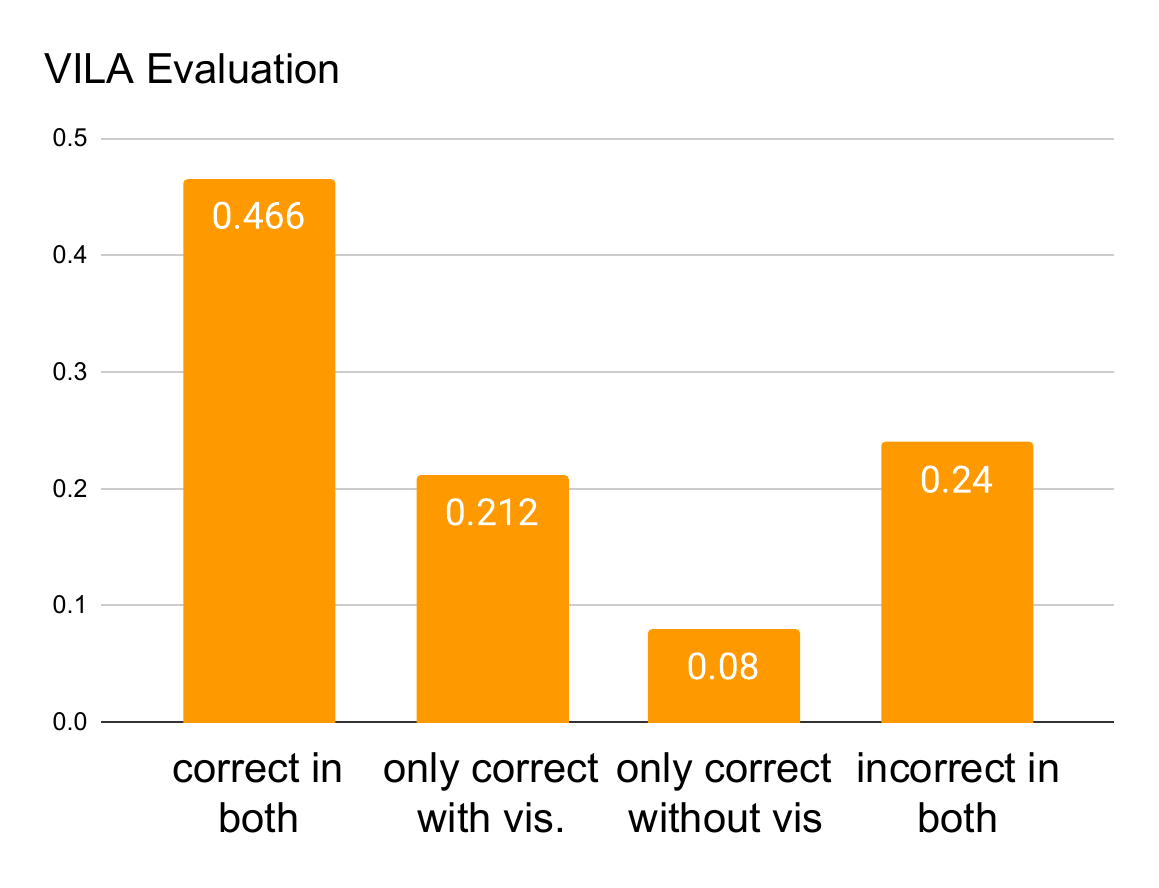}
    \includegraphics[width=0.32\linewidth]{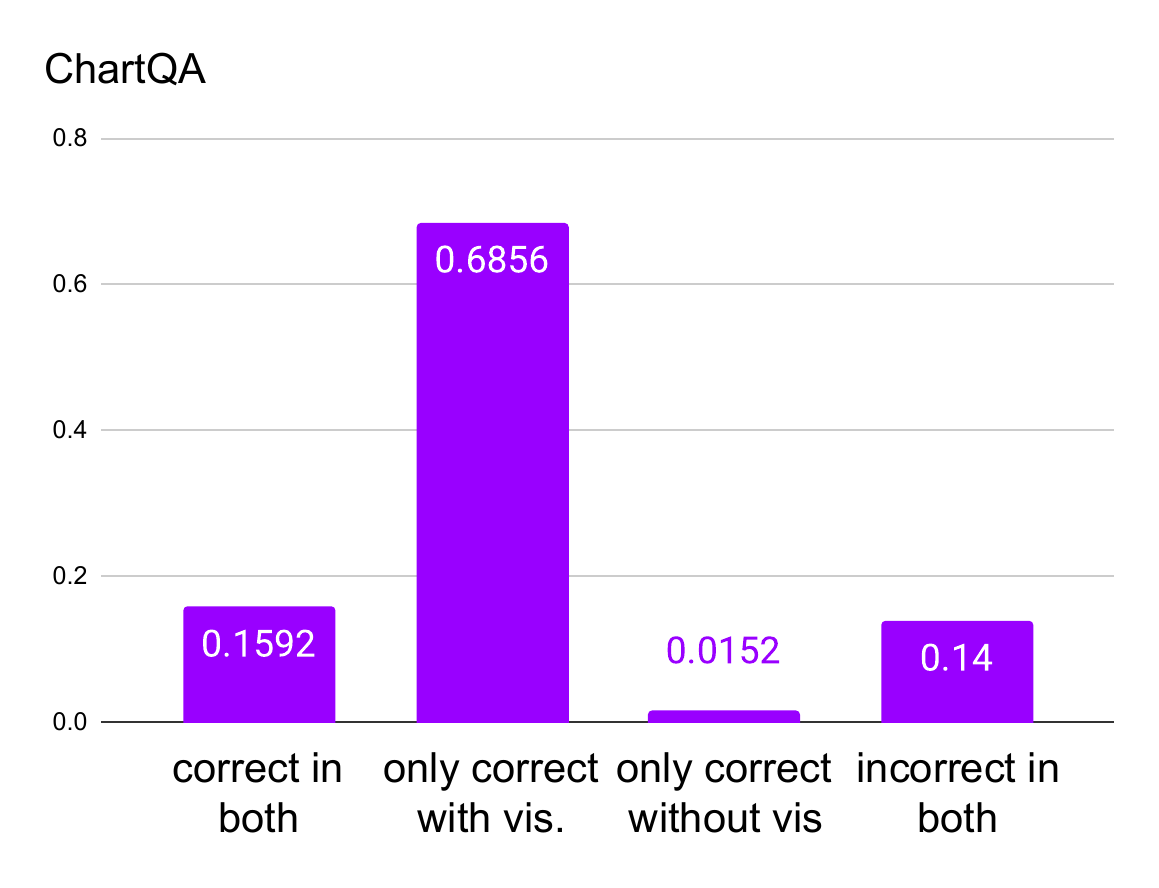}
    \includegraphics[width=0.32\linewidth]{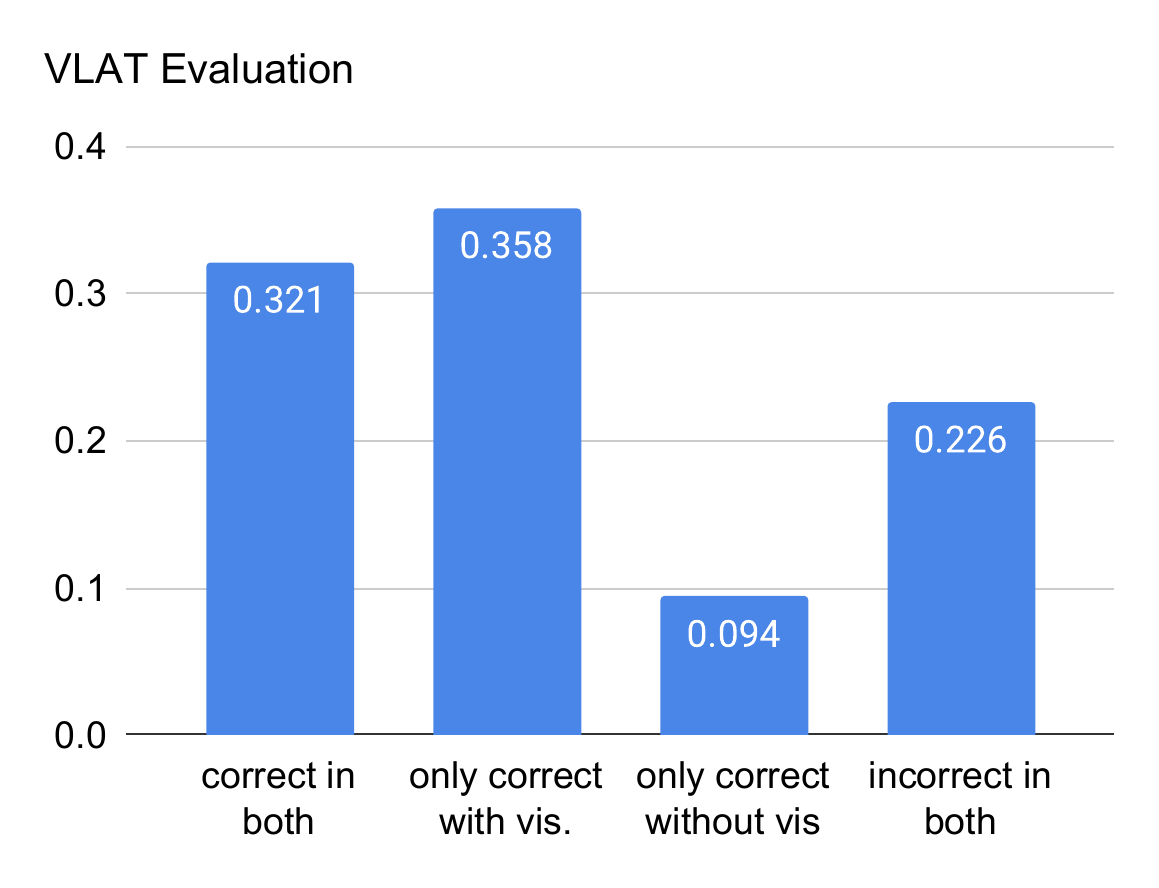}
    
    \caption{The performance of the GPT-4o model with and without visualization during the question query on the VLAT, VILA, and ChartQA datasets. We evaluate these datasets and display the ratio of questions correctly or incorrectly answered in four scenarios. The ratio of questions that are answered correctly in both (with/without visualization) and only answered correctly without vis is high, indicating that there is substantial \textit{recall} in VLAT and VILA evaluation. In the ChartQA, a large portion of questions can only be answered correctly with visualization, revealing a relatively low \textit{recall} rate.}
\label{fig:vlat_ablation_test}
\end{figure*}

\textbf{MLLM and LLM for Visualization}
Our investigation into visualization evaluation practices is motivated by recent work that utilizes both LLMs and MLLMs for a range of visualization problems; please see Yang et al.~\cite{yang2024foundation} for an overview.
Many research efforts have utilized the generative capability of LLMs~\cite{vazquez2024llms, Xu2024ExploringTC, li2024visualization} to create visualizations via code generation, or to interpret and manipulate visualizations through their specifications, e.g., Vega-Lite~\cite{satyanarayan2016vega} or SVG.
Chen et al.~\cite{chen2023beyond} evaluate the performance of GPT3.5 and GPT4 over a data visualization course and find that the LLM can accurately score approximately 80\% of assignments.
Such a finding motivates the new requirement of visualization education design in college. 
In the application of Rich Screen Reader, Zong et al.~\cite{zong2022rich} use LLM to describe visualization to people with impaired vision.
%
Liu et al.~\cite{liu2024ava} proposed the concept of the autonomous visualization agent, which utilizes the visual perception ability of MLLMs to directly understand user intention, adjust visualization parameters, and help solve visualization tasks autonomously. 
The success of these approaches depends on the ability of existing models to reason about visualizations, namely, an understanding of the task presented, and the representation of the visualization, whether expressed in code~\cite{vazquez2024llms} or rendered image~\cite{liu2024ava}.
Our study sheds light on what factors in the input can impact a model's reasoning.
%



\textbf{Machine Perception for Visualization}
Visual perception plays a crucial role in designing an efficient visualization system.
Due to the complexity of human evaluation, many researchers also explore the use of neural network models to assist or even substitute human evaluation.
Haehn et al.~\cite{haehn2018evaluating} examined the performance of convolutional neural networks (CNN) on visualization tasks by testing their ability to handle elementary perceptual tasks as defined by Cleveland and McGill~\cite{cleveland1984graphical}. 
Similarly in this path, Giovannangeli et al.\cite{giovannangeli2020toward} use a neural network to predict human performance on graph-based tasks, provided node-link diagrams or adjacency matrix designs.
Yan et al.~\cite{yang2023can} designed an experiment to measure human behavior on correction comparison in scatter plots with the assistance of different neural network architectures.
More works~\cite{giovannangeli2020toward,zhang2020viscode} are exploring this topic to build the connection between neural networks and human visual perception.
The above works focus on building models specifically for a given visualization task.
This stands in contrast with using MLLMs for visualization reasoning, where models are not optimized for specific visualization objectives, but rather through appropriate prompting, we can use these models to complete a variety of visualization tasks.

\textbf{Evaluation of MLLM Vision Capabilities.}
MLLMs have been broadly accessed on natural images across visual tasks~\cite{bursztynrepresenting} such as visual question answering~\cite{hu2024bliva,wu2024chartinsights,do2023llms,hoque2022chart,chen2024viseval} and image captioning~\cite{bucciarelli2024personalizing}.
Recently, a surge of studies has sought to evaluate MLLMs on chart-based images, such as chart question answering (CQA)~\cite{zeng2024advancing}, chart summarization~\cite{kantharaj2022chart}, and visualization creation~\cite{vazquez2024llms}.
Some of them rely solely on textual representations of images~\cite{tang2023vistext,Xu2024ExploringTC}. 
For example, Bursztyn et al.~\cite{bursztynrepresenting} use chart specifications, which describe the underlying data and visual encodings, to fine-tune and test LLMs on CQA and visual explanation generation tasks.
Other studies directly evaluate MLLMs on images and questions.
For instance, DracoGPT~\cite{wang2024dracogpt} explores visualization design preferences embedded in LLMs via visualization ranking and recommendation tasks. 
Similarly, Alexander et al.~\cite{alexander2024can} and Lo et al.~\cite{lo2024good} demonstrate the potential of MLLMs in detecting misleading visualization; 
while Guo et al.~\cite{guo2024understanding} and Wang et al.~\cite{wang2024aligned} focus on the alignments between MLLMs and humans on graphical perception tasks and chart takeaways predictions, respectively.
The above efforts assume that MLLMs derive answers by interpreting the visualization, overlooking that MLLMs may already possess knowledge about the questions.
A notable exception is the recent work of Endeck et al.~\cite{bendeck2024empirical}, where VLAT is employed as a way to evaluate MLLMs, and ablations are performed to test effects of deceptive design choices, misleading labeling (e.g. titles), and manipulation of data.
Our study differs in that we focus on what factors within CQA are necessary for a model to make a correct decision, for which it turns out the visual encoding itself might play a diminished role.

As MLLMs are trained with a vast corpus of Internet data, it is widely understood~\cite{yuan2024towards} that these models encode a large amount of information.
In the mathematical reasoning~\cite{mirzadeh2024gsm,gulati2024putnam}, researchers found that MLLM may rely on prior knowledge to address the given mathematical reasoning task instead of reasoning through the question.
However, this problem has a limited discussion in the visualization context.
The prior knowledge that may affect the visualization literacy of MLLM has also been discussed in parallel research from Hong et al.~\cite{hong2025llms}.
Their discussion mainly focuses on the VLAT dataset and does not focus on constructing a comprehensive systematic framework for sanity checks to validate an evaluation. 

\section{Problem Statement}
\label{sec:problem}







This work focuses on exploring the essential question of whether MLLMs derive their answer from the visualization when performing the visualization reasoning test \cite{lee2016vlat, nobre2024reading,cui2024promises,masry2022chartqa}.
For a visualization reasoning test, a chart is presented to participants, followed by a question with several options. To answer the question correctly, participants must properly interpret the visualization. 
Our initial exploration of this question across multiple datasets captured something rather unexpected. 

As shown in Figure~\ref{fig:vlat_ablation_test}, where we directly ask an MLLM (GPT-4o in this case) to answer each question of the VLAT~\cite{lee2016vlat}, VILA~\cite{cui2024promises} and ChartQA dataset~\cite{masry2022chartqa}. The evaluation is under two conditions: with and without providing the actual visualization. 
We track the ratio of questions answered correctly or incorrectly in each condition and categorize the outcome into four cases. ``\textit{correct in both}'' shows the model responses to the test correctly with or without visualization. ``\textit{only correct with vis}'' indicates that the model requires the visualization to finish the test correctly. ``\textit{only correct without vis}'' tells that the presence of the visualization during model evaluation fails the test. ``\textit{incorrect in both}'' shows that with and without visualization, the model can not correctly complete the test.

Surprisingly, with 67.9\% correctness, we only observe that 35.8\% of questions in VLAT require visualization to answer correctly in VLAT.
Given the scale of the VLAT dataset contains only 12 visualizations and 53 questions,
we perform a further evaluation on the VILA~\cite{cui2024promises} dataset with 1103 questions and 108 visualizations, and the ChartQA~\cite{masry2022chartqa} dataset with 1250 questions and visualizations.
The experiment with the ChartQA test dataset~\cite{masry2022chartqa} also has a marginal recall effect. Given that 84.5\% correctness, 68.5\% questions require visualization to answer correctly. 
However, the VILA dataset observation shows that only 21.2\% of questions require visualization to correctly complete the test, but the overall correctness ratio is 67.8\%. 
This observation raises concerns about whether the model answers the question by interpreting the visualization as intended or based on the given context of the question, and whether previous evaluations \cite{bendeck2024empirical, li2024litercy} may misinterpret the visualization understanding capabilities of current MLLMs.

\begin{figure}[!t]
\centering   
    \includegraphics[width=\linewidth]{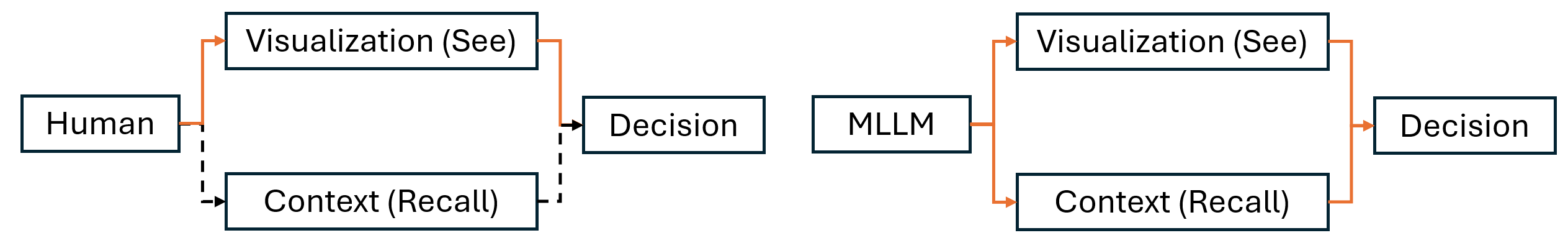}
    \caption{With a presented visualization and a question, the pathway for the decision process of humans and MLLM can be different. Humans often have limited recall knowledge about the asked question and rely on interpreting the visualization to answer the question. In contrast, the MLLM can obtain the correct answer without relying on visualization. 
    }
\label{fig:decision_flow}
\end{figure}

From these initial findings, we hypothesize that the MLLM likely behaves differently from humans, as MLLMs carry vast amounts of information for ``recall'', or the ability to retrieve factual information from the knowledge obtained in their training data, in addition to their ability to ``see'', namely an understanding of visual input that relies on vision capabilities.
To answer a given question, there are two information pathways that an MLLM might follow, as shown in Figure \ref{fig:decision_flow}.
It is possible that the model bases its decision on visualized data, e.g. a value retrieval task would necessitate a model to search, within the visualization, for the data item relevant to the given question.
On the other hand, the model might utilize the context that accompanies the visual encoding of data.
Context can take the form of language input, e.g. question and answer, as well as rendered text, e.g. the title, labels, and axes.
Such context might then be used to trigger recall, or a look-up into the vast knowledge base of the model to answer the question.
We note this is fundamentally different from how a human would reason about the visualization, as such context is rarely sufficient to answer the question without reading the visualization correctly.

However, even if we know the MLLM model can answer many visualization literacy test questions without actually seeing the visualizations, it cannot be concluded that it relies on factual recall to answer the same question when the visualization is provided.
In fact, due to the inaccessible nature of currently available frontier models (e.g., GPT-4o, Claude-3.5), the exact answer may never be possible without direct inspection of the model's internal activations.
Instead, the goal of this work is to understand better how these two pathways can potentially interact with each other, and best estimate the effect of each through a series of carefully designed experiments that ablate various factors and perturb the factual data.
Essentially, we seek to disentangle the impact of \textit{recalling} from that of \textit{seeing} in the model's responses. 
Moreover, by utilizing what we learn from these experiments and observations, we can provide a better assessment of the reliability of existing tests, and then propose guidelines to identify and mitigate the effects of ``recall'' in visualization reasoning evaluation.




\section{Experiment Design}

\begin{figure*}[!t]
\centering   
    \includegraphics[width=\linewidth]{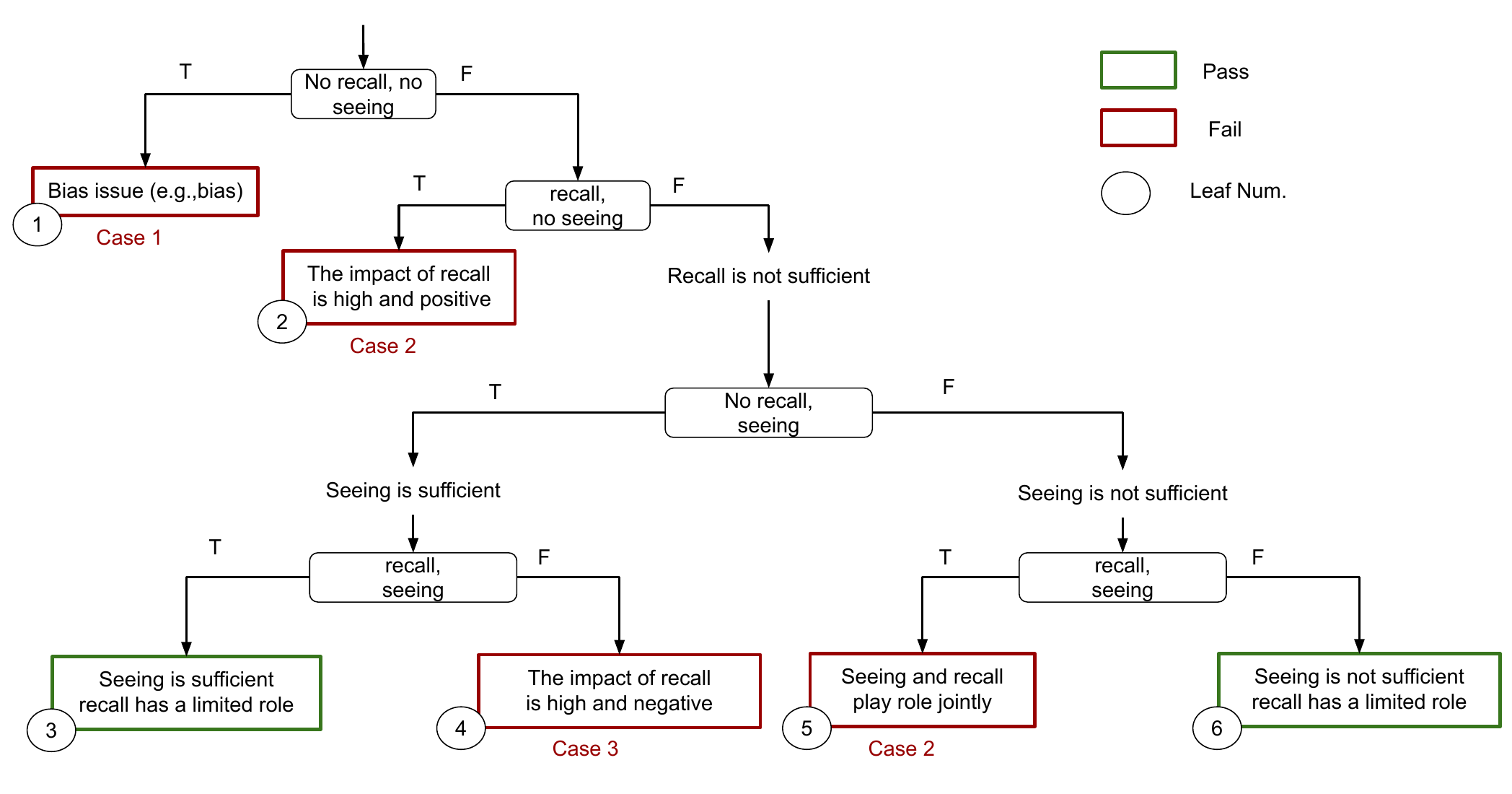}
    \caption{A decision tree-based workflow to identify the problematic cases. The construction of the decision tree begins with no information (no recall/no see) during the visualization reasoning evaluation to full information (recall/see) for evaluation. The construction of the decision tree ends up with 6 leaf nodes, with four nodes failing the sanity check and two passing it. The overall summary led to three cases to validate the model evaluation.}
\label{fig:decision_tree}
\vspace{-2mm}
\end{figure*}

To achieve the goals outlined in Section \ref{sec:problem}, our experiment design needs to untangle the confounding factors and help elucidate the underlying behavior.
As a result, we aim to control and contrast both the source and pathway of information, specifically, 1) what information is encoded in the visual reasoning evaluation (Section \ref{sec:vlatforge}); and 2) how such information is presented to the model (Section \ref{sec:ablation}).


\subsection{Factual Data vs. Non-factual Data}
\label{sec:vlatforge}
Training data for MLLM is often crawled from the Internet, which means an MLLM likely already knows the subject matter and related facts in the evaluation question. 
Our initial experiment (see Figure \ref{fig:vlat_ablation_test}) already validated this suspicion.
As a result, one way to inhibit recall is to modify the data so that it is no longer factual.
To provide independent control over the factual nature of the data, we designed a visualization generator \textbf{VLATForge} (A detailed description of VLATForge can be found in the Appendix) that generates visualization question-answer examples, where the intention is to follow a similar format as VLAT, but with randomly generated non-factual data. 
VLATForge uses the VLAT as a template but fills it with generated data and context. 
This framework is implemented in the Python environment and integrated with the Vega-Altair~\cite{satyanarayan2016vega} library.


\textbf{Factual Baseline}. The factual baseline is the original default setup used to evaluate the performance of MLLM for visualization literacy.
The context and background data used in the VLAT test are from real-world sources.
Most of this information can be found through a Google search, therefore, it is highly likely that an MLLM has come across this information in its training process.

\textbf{Non-factual Randomized Data}. The non-factual data configuration essential retains the same type of visualization and questions, except that the data used for generating the visualization and the corresponding answer are randomly generated matching pairs. The randomness ensures that the correct answer can only be inferred from the visualization alone.  
Such a configuration requires the model to interpret the visualization instead of recalling from prior knowledge, but a human should be able to answer this configuration similarly to the factual version.


Beyond the factual baseline from the VLAT dataset and non-factual dataset generated from VLATForge. Our evaluation also involves a sanity check with two larger datasets.
The Q\&A in the VILA~\cite{cui2024promises} is the same as the VLAT evaluation which gives a question and provides multiple options for selection. 
The evaluation of the ChartQA~\cite{masry2022chartqa} is different in that a question needs a concise answer instead of an option. For example, "How many dollars did dog owners spend on pet food per year in 2020?" and the answer is "442".




\subsection{Information Pathway Ablation}
\label{sec:ablation}
As illustrated in Figure \ref{fig:decision_flow}, MLLMs likely have two information pathways for obtaining answers to visualization literacy questions.
Besides trying to control the data (Section \ref{sec:vlatforge}), we can explore and estimate how each pathway influences the prediction.
We aim to better understand the role of \textit{context} versus \textit{vision} in how a model determines the answer. 
Specifically, context refers to text information that can be used for factual recall, while vision refers to the visual understanding of the visualization itself.
To achieve such a goal we design the following ablations, i.e., the process of removing certain conditions while conducting the same overall experiment, for disentangling and assessing the impact of ``seeing'' and ``recalling''.



\textbf{Remove Visualization (No ``Seeing'')}. 
The visualization contributes to the component of ``seeing'' during the MLLM evaluation.
In this ablation configuration, the plot is removed from the evaluation process, e.g., this removes the information pathway of ``seeing'',  but the rest of the components remain the same. 
If the MLLM can answer the question correctly, it implies that the other elements beyond visualization in the question-answering context are sufficient for the MLLM to make the right decision.
It challenges the assumption that the MLLM actually ``sees'' the visualization to provide the right answer. 
    
    
    
\begin{figure}[!t]
\centering   
    \includegraphics[width=0.45\linewidth]{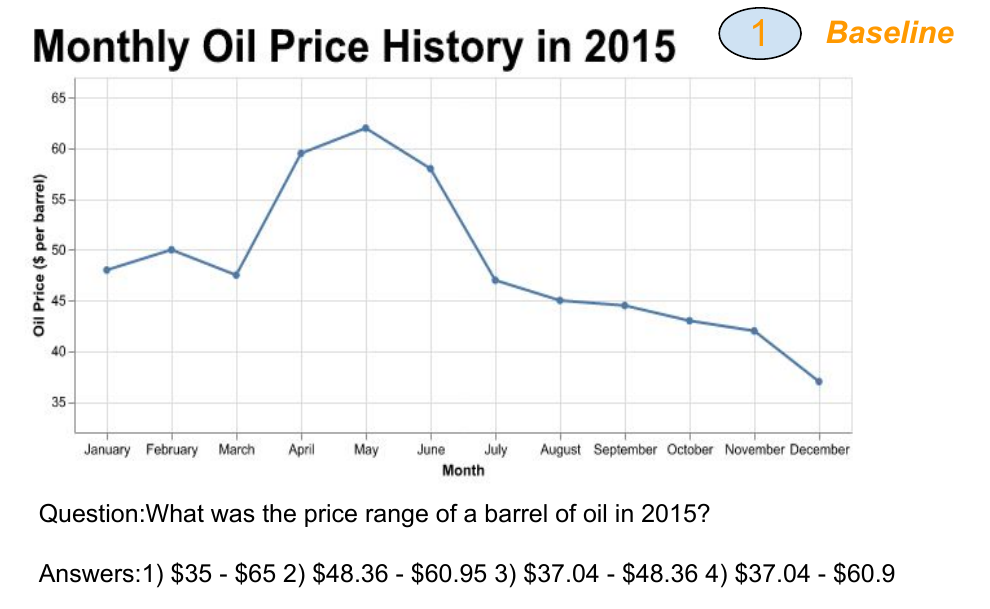}
    \includegraphics[width=0.45\linewidth]{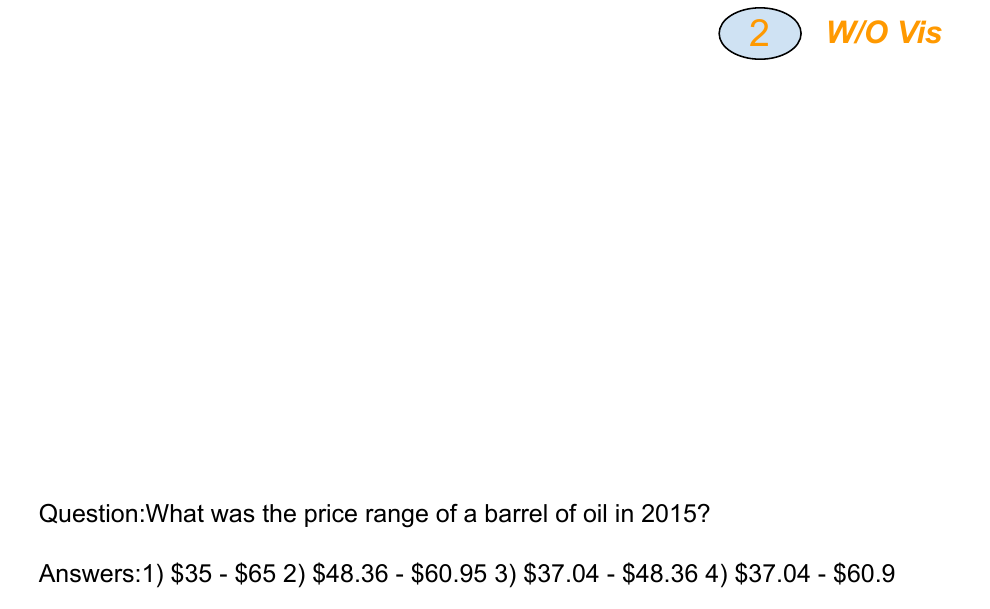}
    \includegraphics[width=0.45\linewidth]{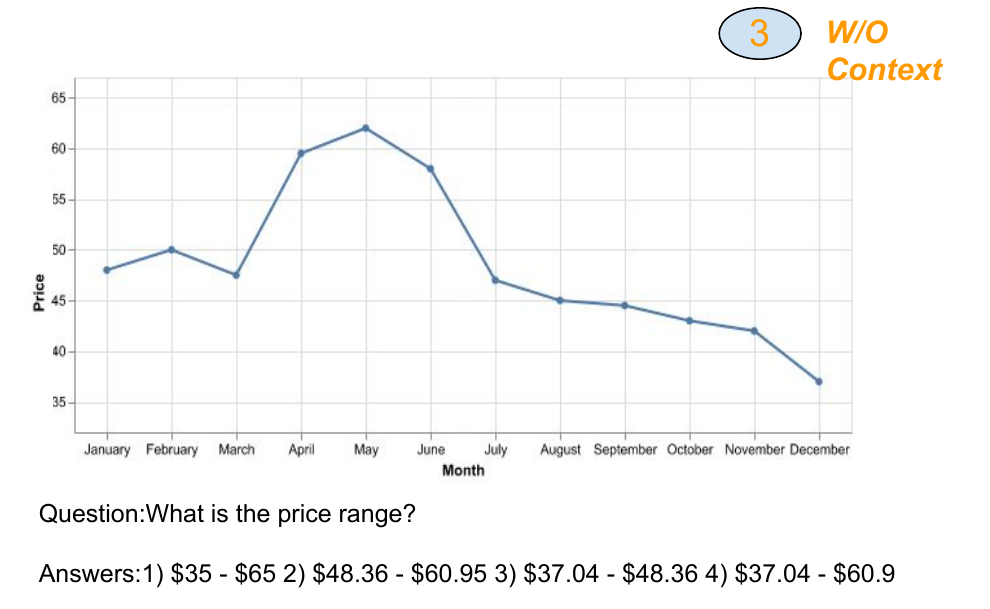}
    \includegraphics[width=0.45\linewidth]{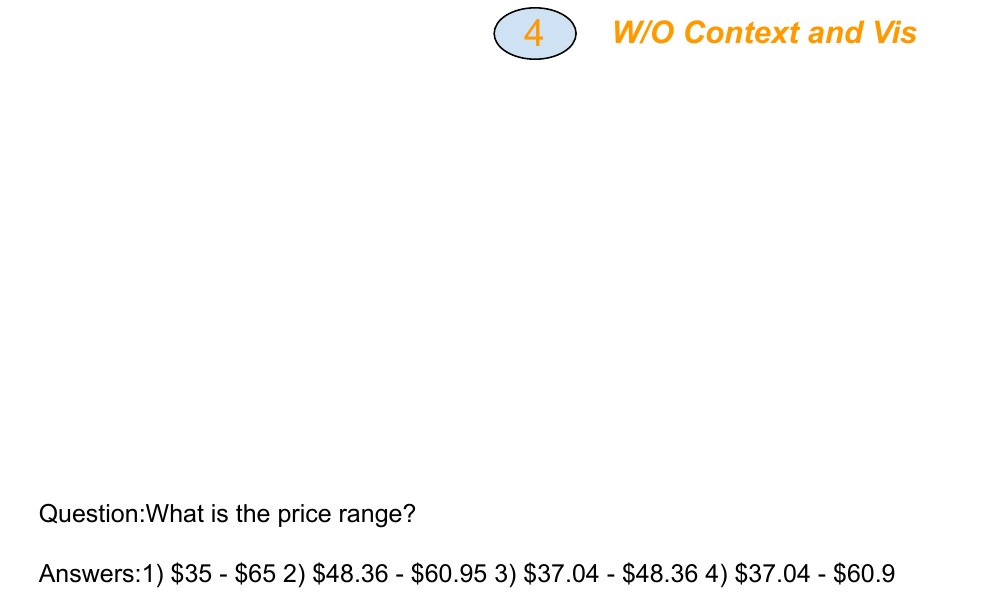}
  
    \caption{Four experiment setups for the four entries of the sanity check table. Similar to the sanity check table, in the 2x2 image grid, the row encodes whether the context is provided, whereas the column specifies whether the visualization is used for evaluation.}
\label{fig:experiment_configuration}
\end{figure}

\textbf{Remove Context (No ``Recall'')}. 
The context refers to any information in the provided question or visualization that would trigger the model to refer to prior knowledge (e.g., year, product, company name, and age group). 
In this ablation configuration, 
we remove all the relevant text in the visualization and question and only preserve the core question referring to the visual elements. 
The \textit{remove context} ablation configuration tries to remove the impact of prior knowledge. 
It helps provide insights into whether the context affects the final decision of the MLLM model positively or negatively.
Combined these two ablation configurations lead to four experiment setups, as illustrated in Figure \ref{fig:experiment_configuration}.

\subsection{Sanity Check Scheme}
Given the ablation options during the evaluation, how to systematically interpret the outcome of the ablation study, and sanity check the evaluation is not trivial. We model the evaluation result as a decision-tree paradigm in Fig.~\ref{fig:decision_tree}. 
We then follow the decision tree diagram to construct a more concise and structured testing setup, coined as the \emph{sanity check table}, that highlights three types of interactions between the vision and the recall information pathway to help reveal potential problems of a given benchmark example.

\textbf{Decision Tree.}
To validate an evaluation, the tree begins with no relevant information (``no recall, no seeing'') presented to the model for testing. 
\textbf{T} means correct and \textbf{F} means Incorrect.
The model should not give a correct answer without any reference information.
If it does \circled{1}, it reveals the inner bias of the model, and we label it as Fail, and the tree will not expand in that branch.
The next step is to validate whether the model completes the test correctly with ``recall, no seeing''.
The correct answer \circled{2} from the model indicates that recall can address the question, and we label it as Fail because recall plays a positive role.
Also, the tree will not expand under this node.
After that, we show ``no recall, seeing'' questions to the model and get a correct answer. If we repeat the experiment with ``recall, seeing'' the final result is correct \circled{3}. 
This passes the sanity check because the presence of the recall does not help the model to answer the question correctly but the visualization will. 
If the result is incorrect \circled{4}, then this is labeled as Fail, because the presence of recall leads to an incorrect answer. However, without recall, the answer is correct, which indicates that recall plays a negative role in the decision.
In the other branch of ``no recall, seeing'' which is incorrect, if the present with ``recall, seeing'' leads to the correct answer \circled{5}, then, it is labeled as Fail because the presence of recall and seeing individually can not lead to the correct answer but jointly. Therefore, recall plays a positive role.
In the end, if the ``recall, seeing'' is incorrect \circled{6}, it passes the sanity check test.

\textbf{Sanity Check Table.}
The final decision tree has six nodes, four of which are invalidated due to model bias or issues with recall. Two nodes pass the sanity check. We can categorize failed nodes into three cases. In case 1, the model has some inner bias. In case 2, recall plays a positive role. In case 3, the recall has a negative role."
The decision tree is a binary decision model. However, given the nature of the evaluation results are measured in probabilities (percentage of success, 
i.e., we run each setup multiple times to account for the inherent stochasticity of the LLM). It is necessary to connect the analysis to a probabilistic perspective.
We introduce a \emph{sanity check} table to summarize these different scenarios in the decision tree from the probability perspective.

\begin{table}[hpbt]
\caption{Sanity Check Table.}
\begin{center}
\begin{tabular}{c|c|c} 
 \hline
  Vis Q\&A   &  Seeing  & No Seeing \\\hline
 Recall    & P1              & P3  \\ 
 No Recall & P2              & P4  \\ \hline
\end{tabular}
\label{tab:sanity_check}
\end{center}
\end{table}

The overall structure of the sanity table is defined in Table~\ref{tab:sanity_check}, but it shares the same configuration as the decision tree.
The row of the table corresponds to the ``recall'' ablation, whereas the column corresponds to the ``see'' ablation. We use P1-P4 to present results from each configuration, and the number indicates the average accuracy of the experiments under that specific condition.
The P1 represents the default baseline configuration, by retaining both ``seeing'' and ``recall''.
The removal of context in visualizations and questions eliminates the possibility of ``recall'' (P2). The removal of visualization highlights the pure recall performance (P3). 
The last case P4, which removes both context and visualization, essentially eliminates both information pathways.

We consider the correct (T) and incorrect (F) as low and high probability.
The high and low in each case are based on the context and the number of experiments performed during the evaluation. 
It can be a threshold number set by the dataset designers.
For example, a binary selection question with  $p>50\%$ correctness can be considered as high probability.



\begin{itemize}
    \item \textbf{Case 1 (C1) model has inner bias}:  P4 represents the impact of other factors beyond our control. It means the model provides a more accurate answer when neither vision nor clear contextual information is presented, indicative of learned biases of the model, e.g., a tendency to revert to the mean, or potentially extrema. In the decision tree Fig.~\ref{fig:decision_tree}, it refers to the leaf node \circled{1}.

     \item \textbf{Case 2 (C2) recall plays a positive role} A high P3 indicates that the impact of ``recall'' is high. The context alone is sufficient for the model to choose the right answer. In Fig.~\ref{fig:decision_tree}, this refers to leaf node \circled{2}. P1 is high but P2, P3, and P4 are low, indicating that present visualization and context individually can not complete the test well. However, their joint impact of them helps the model to answer the question correctly. In Fig.~\ref{fig:decision_tree}, this refers to leaf node \circled{5}.

    \item \textbf{Case 3 (C3) recall plays a negative role}: P1 is low and P2 is high, it can indicate that ``recalling'' provides a strong negative role, e.g., the model mixed up the facts or hallucinated entirely. In Fig.~\ref{fig:decision_tree}, this refers to leaf node \circled{4}.
\end{itemize}

The construction of the decision tree should have 16 nodes because of 4 scenarios.
However, the way we construct the decision tree efficiently reduces the number of cases to six and makes the analysis process easier.
Meanwhile, the construction also implies the priority of the problematic cases.
The verification of an evaluation starts with check Case 1, and if it fails Case 1, the sanity check should reject this evaluation. 
No further verification is needed to justify whether the evaluation is against Case 2 and Case 3.
Otherwise, the evaluation will continue to check whether the evaluation will be checked over Case 2 (\circled{2}, \circled{4}).
If not, then, a further justification is needed to validate Case 3 (\circled{4}). 
If an evaluation passes all three cases, we claim that this evaluation is not affected by the context.

In addition, the sanity check table can be applied to both factual data and non-factual generated data, which can lead to different interpretations.
Moreover, to justify the conclusion of the comparison, it is important to perform a statistical hypothesis test (e.g., a chi-square test or t-test) and report the p-value to justify the significant difference between the distribution of P1 and P2. It is also necessary to statistically distinguish the difference between P3 or P4 with the expected performance under random guessing.

\subsection{Setup and Implementation}
Given that the GPT-4o model outperforms other MLLMs in previous visualization literacy evaluation~\cite{li2024visualization}, and we observe similar behavior for other models, we primarily use GPT-4o all for experiments.
Specifically, we used version gpt-4o-2024-05-13 through OpenAI's API. 
The experiments are performed on the VLAT data with four experiment setups (see Figure \ref{fig:experiment_configuration}) for both factual and non-factual data.
Because of the random seed used during the MLLM execution, the MLLM may not respond to the same question with the same answer in each query. 
To account for the uncertainty, for each question, we recorded 30 different runs. 
A correct answer is scored as 1 an incorrect answer is scored as 0, and the final result is summarized as the average accuracy over all trials.
For example, 15 correct answers over 30 trials is reported as 50\% accuracy.
In this experiment, we use a minimal prompt to query the model and eliminate other potential influences from prompts. The following is an example (for Figure \ref{fig:example}):

"\textit{What was the price range of a barrel of oil in 2015?
Answers: 1) \$35 - \$65 2) \$48.36 - \$60.95 3) \$37.04 - \$48.36 4) \$37.04 - \$60.9\\Why: and please pick one answer in Answer: 1) - 4)}"


\section{Experiment Results and Analysis} 
We start our analysis by closely examining MLLM's response to the VLAT, VLATforge, VILA, and ChartQA datasets aided by the proposed sanity check table. 
We show that the effect of ``recalling'' can greatly influence the evaluation of these datasets. 
With the assistance of the table, we select representative problematic cases as evidence for the impact of ``recalling'' during the evaluation. 
Additionally, we explored the counterfactual analysis with the non-factual benchmark produced by VLATForge, following a similar analysis. 
The experiments query the language model more than ten thousand times involving thousands of different visualizations and questions in this study.

\subsection{A Detail Sanity Check Over VLAT}

\begin{table*}[!htbp]
\setlength{\tabcolsep}{2.5pt} 
\centering
\begin{tabular}{ccccccccccl}
 \hline
Visualization  &   ID   & P1 (See + Recall) & P2 (See + No Recall) & P3 (No See + Recall) & P4 (No Recall + No See) & C1 & C2 & C3\\  
\hline
{}                 & Item 1 & 0.2& 0.1& 0.03& 0.2&- &-&-\\
{}             & Item 2  & 1.0& 0.96& 0& 0.03&- & - & -\\
Line Chart (1) & Item 3  & 0.666& 0.466& 0.966& 0.9 &  \checkmark &  - & - \\
{}             & Item 4  & 1.0& 1.0& 0.666& 0.33&- &  \checkmark & -\\
\hline
{}                 & Item1 & 0.6& 0.233& 0.2& 0.133& - & \checkmark & -\\
Bar Chart(2)   & Item2 & 1.0& 1.0& 1.0& 0.533&-& \checkmark &-\\
{}                 & Item3 & 0.133& 0.2& 0.166& 0.06&-&-&-\\
{}                 & Item4 & 0.533& 0.433& 0.33& 0.3&-&-&-\\
\hline
{}                 & Item1  & 0.233&        0.433&                0.266& 0.433& - & - & \-\\
Area Chart(8)  & Item2 & 0.03&        0.93&                0& 0& - & - & \checkmark \\
{}                 & Item3 & 0.5&        0.56&                0.3& 0.36&-&- & -\\
{}                 & Item4 & 1.0&        0.93&                0.66& 0.13&- &  \checkmark & -\\
\hline
{}             & Item 1  & 1.0&        0.43&                1.0& 0.3&  - &  \checkmark &  -\\
Treemap(12)    & Item 2  & 1.0&        1.0&                0.26& 0.0& - & - &  -\\
{}             & Item 3  & 1.0&        1.0&                1.0& 1.0&  \checkmark &  - & -\\
 \hline
VLAT dataset &  & 0.576 & 0.552 & 0.427 & 0.374 & - & - & - 

\end{tabular}
\caption{The (partial) table displays the sanity check of MLLM over each question of the VLAT datasets (see the full table in the appendix). The problematic cases are marked in the last three columns C1-C3 provided the observation is statistically significant.}
\label{tab:vlat_question_data_table_sanity_check}
\normalsize
\end{table*}

We use the VLAT benchmark to demonstrate how our design framework performs the sanity check and highlights the problem of each question. 
Despite multiple attempts \cite{li2024litercy, bendeck2024empirical,hong2025llms} to leverage VLAT for evaluating MLLM's visualization understanding capabilities, limited works investigated deep in the relationship of vision vs. factual recall in their evaluation. How the evaluation is affected by the intended bias of the model and context recall.
The partial result of the sanity check is presented in Table~\ref{tab:vlat_question_data_table_sanity_check} (please refer to the full version in the appendix), where the middle columns specify the configuration of the experiment, and the last three columns check whether a given question belongs to any problematic cases that we identified (see Section \ref{sec:ablation}).
Given that each question is performed with the model 30 times, we use the chi-square test to analyze whether the result is statistically significant.
The cell in the table for each case is only checked when $p < .05$. 

From the sanity check table of the entire VLAT dataset (summarized across 30 repeated runs), a majority of the questions seem to rely on recall. 
The numbers in the table represent the mean accuracy of all runs under the respective configuration.
For example, in Table~\ref{tab:vlat_question_data_table_sanity_check} bar chart item 1, under the \emph{see + recall}, the model responds with a 60\% accuracy.
\emph{see + no recall} has 23.3\%, \emph{no see + recall} has 20\%, and \emph{no recall + no see} has 13.3\% performance.
Given that the performance between seeing and without seeing shows a significant difference, we can tell that the evaluation is case 2, and the context playsa  positive role during the evaluation.

\subsubsection{Inherent Bias Beyond Context (Case 1)}
The model may possess inherent biases influencing its answer selection, independent of any visual or contextual information.
For example, consider Line Chart Question 1, Item 3, as illustrated in Figure~\ref{fig:experiment_configuration}. As shown in Table~\ref{tab:vlat_question_data_table_sanity_check}, the model attained a remarkable 90\% accuracy under the P4 condition, where both visualization and context were absent. When inspecting the model's responses, it appears that the model is making decisions based solely on analyzing the provided answer options.
The question and answer options without context are as follows:

\textit{Line Chart (1) Item 3 (no context): What was the value range?
Answers:1) 35 - 65 2) 48.36 - 60.95 3) 37.04 - 48.36 \textbf{4) 37.04 - 60.95}}

In this scenario, the model consistently selects Option 4. The GPT-4's response provides the following explanations:

\textit{\textbf{GPT-4o Response}: ``Answer: 4) 37.04 - 60.95 This range combines the two overlapping ranges from options 2) and 3) to cover the entire spread''.}

This indicates that the model is utilizing reasoning based on the numerical patterns and relationships among the answer choices themselves, rather than relying on any visual or contextual cues. The model seems to favor options that represent a comprehensive range or an aggregate of other options, suggesting an inherent bias toward certain answer structures.
This finding raises important considerations about the model's behavior in the absence of external information. The ability to generate correct answers based solely on the answer options implies that the model may exploit inherent biases or heuristics. Such biases could stem from patterns learned during training, where certain numerical relationships or answer formats are more commonly associated with correct answers.

Understanding this behavior is crucial, as it highlights another dimension where the model's performance may be overestimated if these inherent biases are not accounted for. It underscores the need for carefully designed evaluation questions that minimize the possibility of the model leveraging such biases. By ensuring that answer options do not inadvertently provide clues or patterns that the model can exploit, we can achieve a more accurate assessment of the model's true capabilities in interpreting visualizations.

\subsubsection{Context Plays Positive Role During Evaluation (Case 2)}
\label{sec:case-1}
In our analysis of Table \ref{tab:vlat_question_data_table_sanity_check}, we found that for many of the questions, the model's performance aligned with Case 2, where ``recalling'' has positively influenced the outcomes.  
To understand why this behavior occurred, we examined Question 12 as a representative example. 

\textit{Tree Map (12) Item 1: which website was the number of unique visitors the largest in 2010?
Answers:1) Facebook 2) Amazon 3) Bing 4) \textbf{Google}}

\textit{Tree Map (12) Item 1(no context):  which item has the largest portion in the visualization? Answers:1) D1 2) B1 3) A2 4) \textbf{A1}}
            
\begin{figure}[!htbp]
 \centering   
     \includegraphics[width=\linewidth]{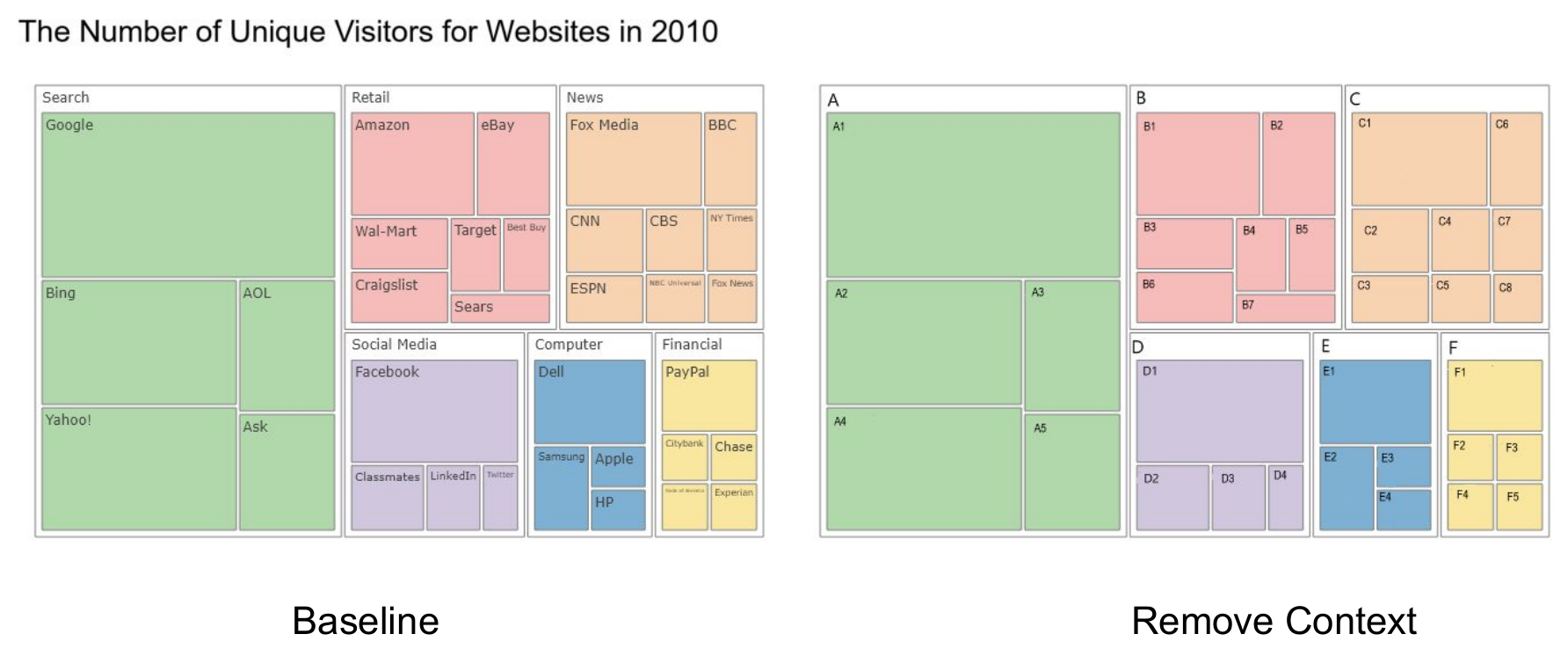}
     \caption{In the treemap, the left visualization is used as the see + recall for MLLM evaluation, and the right visualization display is the case for no recall evaluation.}
\label{fig:treemap_baseline_no_context}
\end{figure}

This question involves interpreting the treemap evaluations in Fig.~\ref{fig:treemap_baseline_no_context}. The task involves measuring the model's accuracy in interpreting a treemap where the size of each rectangle represents the number of visitors to different company websites in a specific year. In the original treemap, company names are displayed, while in the modified version without context, company names are replaced with generic alphabetic characters.

Analyzing the sanity check for Question 12, Item 1 in Table~\ref{tab:vlat_question_data_table_sanity_check}, we observe that the P2 value shows an accuracy of 43\%, whereas the P1 value is at 100\%, indicating that removing the visualization significantly increases the model's performance. This is further supported, by the P3 value also displays 100\% accuracy, confirming that the model can answer these questions correctly without any visual input, relying solely on contextual recall.
To statistically validate both observations, we conducted a hypothesis test comparing the P1 (\emph{See + Recall}) and P2 (\emph{See + No Recall}) values. The difference in performance was statistically significant ($p < .001$), indicating that recall has a significant influence over the model's accuracy in these tasks.
These observations demonstrate that context significantly enhances the model's accuracy, and the overall evaluation of these questions may be overestimated if the impact of recall is not properly accounted for. This finding is evidence for the recall issue, as it reveals that the model may perform significantly better.

\textbf{Context Removal Difficulty}. To make the Case 2 and 3 assessment, we assume that context can be removed in most visualizations, however, in some cases, fully removing the context is not possible. For example, in Choropleth maps, where the visual representation itself already encodes context, e.g. the shape of a state offers already contextual information that cannot be removed by merely removing the text labels.

\subsubsection{Context Plays Negative Role in Counterfactual with Non-Factual Data (Case 3)}
As previously demonstrated, the context of the VLAT dataset is factual and improves the evaluation accuracy. A natural question to ask is how would the MLLM behave if the original data used for generating the visualization test were non-factual. Would the mismatch between the non-factual data and factual recall from the context create conflict and confusion for the model that decreases their performance and leads to lower overall accuracy? 

\begin{figure}[b]
\centering   
    \includegraphics[width=\linewidth]{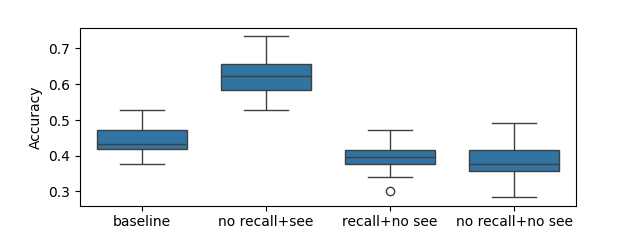}
    \caption{The score distribution of sanity check over the non-factual dataset.}
\label{fig:vlatforge_sanity_check_validate}
\end{figure}

As discussed in Section \ref{sec:vlatforge}, we obtain the non-factual benchmark through the proposed VLATForge tool. 
We assume that with no prior knowledge to draw from, the model has to rely on its vision capability and interpret the visual information to answer the question from the non-factual dataset correctly.
From the sanity Table~\ref{tab:vlatforge_sanity_check} of the overall VLATForge dataset,  we can see the default setup, i.e., \emph{See + Recall}, where both information pathways are preserved, leads to an accuracy of around $P1=42.4\%$. It is a significant drop from the original VLAT baseline with factual data of $P1=57.6\%$ and their difference is statistically different through t-test.
Interestingly, if we remove the context, i.e., \emph{See + No recall}, we can see a much-improved accuracy of $P2=62.8\%$. 
We can also check the score distribution of each case in the experiment from Figure~\ref{fig:vlatforge_sanity_check_validate}. 
The accuracy of the \emph{See + Recall} slightly edges out against the \emph{No See} cases.
The gap between \emph{See + No Recall} and \emph{See + Recall} indicates it belongs to case 3.
The statistic also shows that the context has a negative impact during the model evaluation.
These summary statistical results point to the most likely scenario that MLLMs indeed utilized both information pathways. In this case, non-factual data in the visualization, as well as potential factual recall from the context, lead to conflicting information that reduces the accuracy of the non-factual data VLAT benchmark.

\begin{figure}[hpbt!]
\centering   \includegraphics[width=\linewidth]{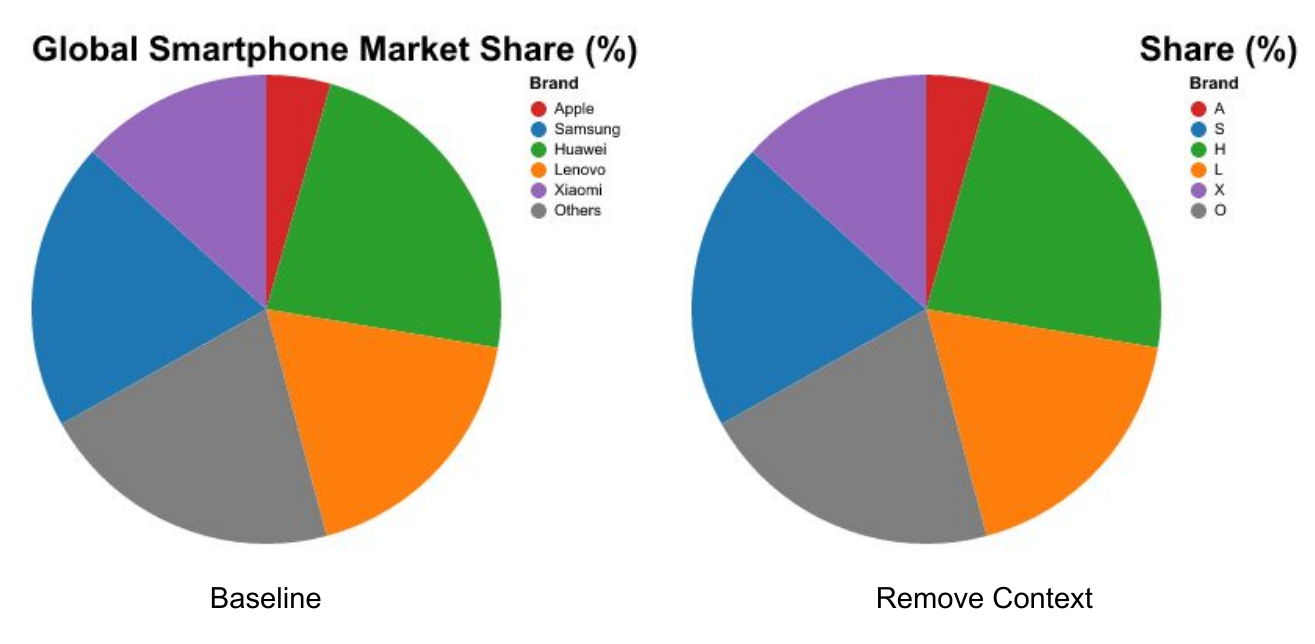}
    \caption{In the area chat, the left visualization is used as the see + recall for MLLM evaluation, and we replace the names of companies with an alphabet character in the right visualization.}
\label{fig:pie_chart_no_context}
\end{figure}

\begin{table}[t]
\begin{center}
\caption{The sanity check over the Bubble Chart from the VLATForge no-factual dataset.}
\begin{tabular}{c|c|c} 
 \hline
  Bubble Chart (10)      & See       & No See \\\hline
 Recall     & 0.3           & 0.0  \\ 
 No Recall  & 1.0         & 0.53 \\ 
 \hline
\end{tabular}
\label{tab:vlatforge_bubble_sanity_check}
\end{center}
\end{table}

To dive deeper into the problem, let's take a closer look at the individual examples.
With the non-factual data, the impact of recalling can be different compared with the previous factual data we see previously, e.g., for the pie chart, the impact of recalling is negative as shown in Table~\ref{tab:vlatforge_bubble_sanity_check}.
The accuracy of the model with \emph{See + Recall} ends up with 30\% accuracy.
In the meantime, if we mitigate the impact of context and use the visualization to decide the answer, the model can make all correct decisions.
On the other hand, if we remove the visualization but keep the context, the accuracy of the model drops to 0\%, as it likely just recalling the factual information even though the correct answer for the non-factual visualization is different.
The question of the evaluation with and without context is framed in the following:

\textit{In which company is the global smartphone market share the smallest? Answers:\textbf{1) Apple} 2) Huawei 3) Samsung 4) Lenovo}

\textit{In which of the following shares the smallest? Answers: \textbf{1) A} 2) H 3) S 4) L }

\textbf{GPT-4o  response}: 
\textit{`` According to the pie chart, Lenovo has the smallest segment, indicating the smallest share in the global smartphone market among the listed companies. Answer: 4) Lenovo''}

The model's responses show that it generally believes that Lenovo is the smallest segment which is the factual knowledge, but the actual case in the visualization is Apple.
Essentially, the VLATForge benchmark with non-factual data explores the counterfactual aspect of context, where we see a clear influence of context in the negative direction.
From the above analysis, we can conclude that the latest MLLM model like the GPT-4o model can be negatively affected by the effect of ``recall'' in visual questions and answering types of evaluations. 

\begin{table}[t]
\begin{center}
\caption{The sanity check over the entire non-factual dataset. The statistic tells us that the evaluation of the non-factual dataset is affected by the ``recalling'' in a negative manner.}
\begin{tabular}{c|c|c} 
 \hline
  VLATForge Dataset       & See       & No See \\\hline
 Recall     & 0.442           & 0.393  \\ 
 No Recall & 0.624           & 0.385  \\ 
 \hline
\end{tabular}
\label{tab:vlatforge_sanity_check}
\end{center}
\end{table}

\subsection{The High ``Recalling'' Rate of the VILA Dataset with LLM Generated Context and Selection}
VILA~\cite{cui2024promises} is a dataset that uses 12 chart types from VLAT as a template, it is intently designed for humans' visualization literacy test by expanding the VLAT to a larger corpus.
During the dataset design, a large language model constructs the question and four selection options of the evaluation automatically, and the visualization data is generated randomly. The visualization has a different color setup and number of data items. 
Visual experts go through each question for quality check and the final dataset is evaluated over a large number of human candidates for visualization literacy analysis.
In Fig.~\ref{fig:vlat_ablation_test}, we have demonstrated that VILA is highly affected by recall.
Interestingly, the model can answer most of the description questions without visualization in this dataset, for example, what the visualization tries to describe or what is the topic of visualization (Detailed examples are provided in the appendix).

To perform a detailed sanity check analysis, we probe the dataset by sampling 50 images and questions uniformly to examine the dataset's quality under our analysis framework.
During the sample selection, we filtered out description-type questions which have a high recall.
The aggregation result of the evaluation is presented in Table~\ref{tab:ChartQA_sanity_check}.
The P4 is the case of ``no see'' and ``no recall'' which has 32\% accuracy which is not significantly different from the random guess baseline. The evaluation passes case 1.
After filtering out the description type of question, we can tell from the sanity check table that ``recall'' and ``no see'' or the opposite have only 34\%. 
The recall or vision itself can not help to answer the question correctly.
However, if both channels are presented, the performance increases to 52\% revealing that vision and recall jointly contribute to the final decision. 
The overall evaluation does not pass case 2 and further refinement of the data is necessary before applying it for MLLM visual reasoning testing.

The result of the evaluation in Fig.~\ref{fig:vlat_ablation_test} and Table~\ref{tab:VILA_sanity_check} show that the evaluation of VILA with such a high recall is a surprise because the data of the visualization is generated randomly. 
It raises concerns about whether the context and selection generated by the MLLM may trigger high recalling rate. A further experiment is required to justify the hypothesis, but it is out of the scope of this sanity check study.

\begin{table}[t]
\begin{center}
\caption{The sanity check of the VILA dataset}
\begin{tabular}{c|c|c} 
 \hline
  ChartQA(subset) Dataset       & See       & No See \\\hline
 Recall     & 0.52           & 0.34  \\ 
 No Recall & 0.34         & 0.32  \\ 
 \hline
\end{tabular}
\label{tab:VILA_sanity_check}
\end{center}
\end{table}

\subsection{Sanity Check Over Concise Response Q\&A in ChartQA Dataset}
The evaluation of ChartQA~\cite{masry2022chartqa} dataset is different from the VLAT and VILA datasets, which requires answering directly without providing multiple options. We use 1250 questions that are generated from the QA from human-written summaries.
Bar, Line, and Pie charts are the major types of visualizations in ChartQA.
An example of the prompt used for evaluation is constructed as:
\textit{"What was the growth rate of real wages in Slovakia in 2017? Please answer the question in the format: Answer:"}.

The result from Fig.~\ref{fig:vlat_ablation_test} which presented with and without visualization reveals 68.5\% of questions require visualization to complete the test successfully and the model answers around 16\% of the questions without visualization.
Compared with the multiple-selection setup in VILA, the overall evaluation of ChartQA seems less affected by the recall during the model prediction.
Similar to the previous analysis, we randomly sample 50 images and questions and use our sanity check framework to examine the quality of the dataset for further examination.
The ``no see'' and ``no recall'' cases have 0\% accuracy which pass the Case 1 check.
Without seeing the visualization, the prediction accuracy is 14\% which is relatively low and passes Case 2 validation.
The last step checks the ``see'' and ``recall'' with ``see'' and ``no recall'', we can tell that the difference between these two categories is subtle.
The observation indicates that the evaluation of ChartQA is less affected by the recalling compared with VILA.

The evaluation accuracy of the model over ChartQA is much higher compared with the VLAT and VILA benchmarks. 
The main contribution may result from the data annotation in the visualization. Previous work~\cite{li2024litercy} has demonstrated that properly annotating the visualization can improve the model's performance over the VisQA. 
A follow-up question would be whether a proper data annotation helps to mitigate the impact of recalling. 
If the impact of recalling positively contributes to the final decision, the data annotation will not help the evaluation pass the sanity check case 2. Similarly, combining visualization with the source data~\cite{bendeck2024empirical} to do the evaluation will have the same problem.
However, if the impact of the recalling contributes negatively to the evaluation, then the annotation or attaching of the source data may help to mitigate the problem.

\begin{table}[t]
\begin{center}
\caption{The sanity check of the ChartQA dataset}
\begin{tabular}{c|c|c} 
 \hline
  ChartQA(subset) Dataset       & See       & No See \\\hline
 Recall     & 0.92           & 0.14  \\ 
 No Recall & 0.86         & 0.0  \\ 
 \hline
\end{tabular}
\label{tab:ChartQA_sanity_check}
\end{center}
\end{table}

\subsection{Can Prompt Redirect the MLLMs' Attention?}
\label{sec:prompt_op}
With the above observation, one easy fix for the potential problem is to request the MLLM (e.g., GPT-4o) to focus on the visualization to answer the design question.
To verify whether we can propose a better prompt that helps the model focus on the visualization to address the challenge, we experiment on the non-factual dataset to explore the effectiveness of the alternative prompt.
Given the impact of context negatively impacts the evaluation, we expected that the prompt that requests the model to focus on visualization to extract information should perform better than the original prompt version.

\begin{figure}[t]
\centering   
    \includegraphics[width=\linewidth]{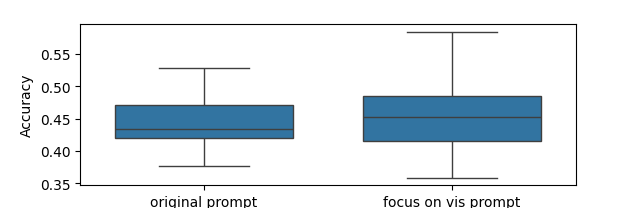}
    \caption{The score distribution of two prompts over the non-factual randomized data.}
\label{fig:prompt_comparison}
\end{figure}

Similar to the previous experiment, we use the following prompt which contains key instruction  \textit{``focus on the information from the visualization to answer the question''}. The experiment is also performed with 30 trials and compared with the experiment with the original prompts.  From the evaluation result Figure~\ref{fig:prompt_comparison}, we can tell that the two experiments' distribution scores do not differ much.

\textit{\textbf{New prompt}: What was the price range of a barrel of oil in 2015?
Answers: 1) \$35 - \$65 2) \$48.36 - \$60.95 3) \$37.04 - \$48.36 4) \$37.04 - \$60.9 
Why: and \textbf{please focus on the information from the visualization} to answer the question and pick one answer in Answer: 1) - 4)}

We further performed a t-test to measure the difference between the score distribution of all questions. The result, the p-value of 0.395, indicates that we can reject the hypothesis that the two distributions are similar. In the alternative, there is no significant difference between the new prompt and the original prompt.
Our experiment indicates that direct prompting may not be a straightforward solution to address the current problem.

\section{Sanity Check-Guided Mitigation Strategies}

Based on our findings, we propose several mitigation strategies to reduce the influence of recall and enhance the reliability of visualization understanding assessments.

\textbf{Understanding the Prevalence of Recall}. The first line of defense against the recall problem is to quantify its prevalence within the evaluation. Utilizing the sanity check table introduced in our study, the researchers can systematically identify the extent to which recall affects the model's performance on individual questions and across the dataset. 
This approach enables researchers to flag questions or entire datasets where recall significantly skews the results. Identifying these instances is essential for understanding the limitations of the evaluation and for guiding subsequent mitigation efforts. 

\textbf{Avoiding Questions with Factual Data}. Another effective mitigation strategy is to minimize the use of factual data that the model might have encountered during training. By employing non-factual or randomly generated data, as implemented in the VLATForge dataset, we reduce the likelihood that the model can rely on prior knowledge to answer questions. However, this can lead to underestimation due to conflicting information from vision and recall. Therefore, the preferred scheme is to remove or alter contextual cues (such as text labels) that might trigger recall. For instance, replacing specific names, dates, or locations with abstract or fictional equivalents can help prevent the model from drawing on prior knowledge. 

\textbf{Designing Counterfactual Evaluation Scenarios}. Creating counterfactual evaluation scenarios is another approach to mitigate recall. By intentionally designing questions and visualizations that contradict common knowledge or the model's likely training data, researchers can test the model's ability to prioritize visual input over prior knowledge. If the model correctly interprets the counterfactual visualization despite conflicting with its training data, it indicates a stronger reliance on visual reasoning.

\textbf{Monitoring and Updating of Evaluation Datasets}. 
Given the rapid progress in the development of MLLMs and their training data, it is important to regularly monitor and update evaluation datasets. As models evolve and are trained on increasingly large and diverse datasets, previously non-factual or synthetic data might become part of their knowledge base. Continual assessment of the evaluation content's uniqueness and its potential inclusion in training data is necessary to maintain the integrity of the evaluations. Additionally, the randomized data generation process proposed in VLATForge can actively combat this challenge as every set of benchmarks it generates will be different.

\textbf{Prompts Engineering to Focus on Visualization}.
Enhancing prompt design to explicitly direct the model's attention to the visual input could reduce recall. 
Even though our initial exploration of the instructional prompt is unsuccessful, this does not mean this can not be a viable path, as more sophisticated prompting methods are used or the model's instruction-following capabilities increase over time.

With the goal of sanity checks in mind, through carefully designed experiments and statistical tests, we have identified several surprising behaviors regarding how contextual information can positively or negatively impact the model's performance in visualization understanding tasks, which shed light on the role of vision in the decision-making process of MLLMs. 
Yet, despite our best efforts to disentangle the effects of ``recalling'' and ``seeing'', the exact answer can still be elusive. For example, some confounding factors are likely impossible to unravel, and the close source nature of the model weights makes it impossible to directly examine activation and attention in the model. 
Therefore, rather than claim we can tell exactly what happens in the decision-making process, we focus on identifying what is certainly not right.
Our study highlights the broader issues of evaluating the black-box nature of these models to reliably assess their true capabilities.

Beyond the more philosophical discussion on what is possible, on the more practical side, there are several limitations regarding the proposed approaches.
First, our work assumes that context can be modified or removed, however as discussed in Section \ref{sec:ablation}, this may be not possible if the context is embedded in the visualization itself such as the choropleth map. 
Second, with the sanity check table, aggregating results can obscure detailed effects observed at individual questions or experimental run level could potentially lead to missing information. For example, the context may have a diverging impact on the accuracy of different questions, even though the summary statistics are similar. 
Third, our experiments are limited to GPT-4o. Despite similar behavior from different MLLM models, direct comparisons may reveal their unique characteristics.

For the future direction, we plan to extend the same sanity check to other models, such as Gemini~\cite{team2023gemini} and Claude.
In addition, we believe the influence of ``recall'' in MLLM evaluations exists beyond the visualization literacy test, many state-of-the-art general visual question-and-answer benchmarks on plots and charts \cite{hu2024bliva, xu2023chartbench, li2024seed, kahou2017figureqa, masry2022chartqa, mathew2022infographicvqa} that are prevalent in machine learning communities also exhibits similar challenges. 
In the future, we aim to expand our evaluations to include a wider range of models and datasets. We will develop automated methods to modify context and visual elements without compromising the integrity of the evaluation tasks is essential. Applying advanced NLP techniques and leveraging adversarial examples could enhance the robustness of evaluation frameworks. Finally, investigating the impact of recall in other modalities and developing cross-modal evaluation strategies could provide a more comprehensive understanding of MLLMs' capabilities. This also opens up opportunities for visualization research to develop new tools that help in analyzing the accuracy of different experimental conditions. 

\section{Conclusion}
Evaluating the visual capabilities of MLLMs is foundational for their effective application in visualization understanding tasks. 
In this paper, we design a comprehensive sanity check framework that critically examines the classical question-and-answer approach used to assess MLLMs' visualization reasoning ability, revealing a critical issue: the potential for models to rely on contextual recall rather than genuine interpretation of visual signals. 
Using the four disparate representative datasets, VILA, ChartQA, VLAT, and VLATForge, we demonstrated that the presence of contextual information may substantially influence the evaluation outcomes, leading to either overestimation or underestimation of the proper visualization understanding capabilities of the MLLMs.
We disentangle the effects of ``recalling'' and ``seeing'' during evaluation and exhaustively go through all combination cases with a decision tree structure to validate a VisQA question.
Building on the decision tree, we introduced a generable sanity check table and three distinct validation cases to justify the context's impact.

Our findings show that the context can obscure the true extent of an MLLM's ability to interpret visual information. 
Such observations lead to questioning the validity of conclusions in previous studies that may have skewed performance metrics.
Furthermore, it raises a broader concern about the generalizability of evaluation practices beyond visualization understanding. 
Similar recall issues may also be prevalent in assessments involving other modalities, such as video and audio.
Ultimately, as these models' visualization understanding becomes increasingly more capable, it becomes even more important that we conduct rigorous and nuanced evaluations to ensure valid evaluation of their capabilities. 
In an era where artificial intelligence is becoming increasingly popular, our work underscores the value of a more rigorous evaluation of these models. 
By discovering the challenges associated with recall and proposing mitigation strategies, we contribute to the foundation of trustworthy AI, ensuring that these increasingly capable models can be relied upon to perform tasks accurately. 

\acknowledgments{
The authors wish to thank A, B, and C. This work was supported in part by
a grant from XYZ.}

\bibliographystyle{abbrv-doi}

\bibliography{main}
\newpage

\begin{appendices}

\section{The ``Recalling'' Problem in the  Claude Model}

\begin{figure}[!hpbt]
\centering   
    \includegraphics[width=\linewidth]{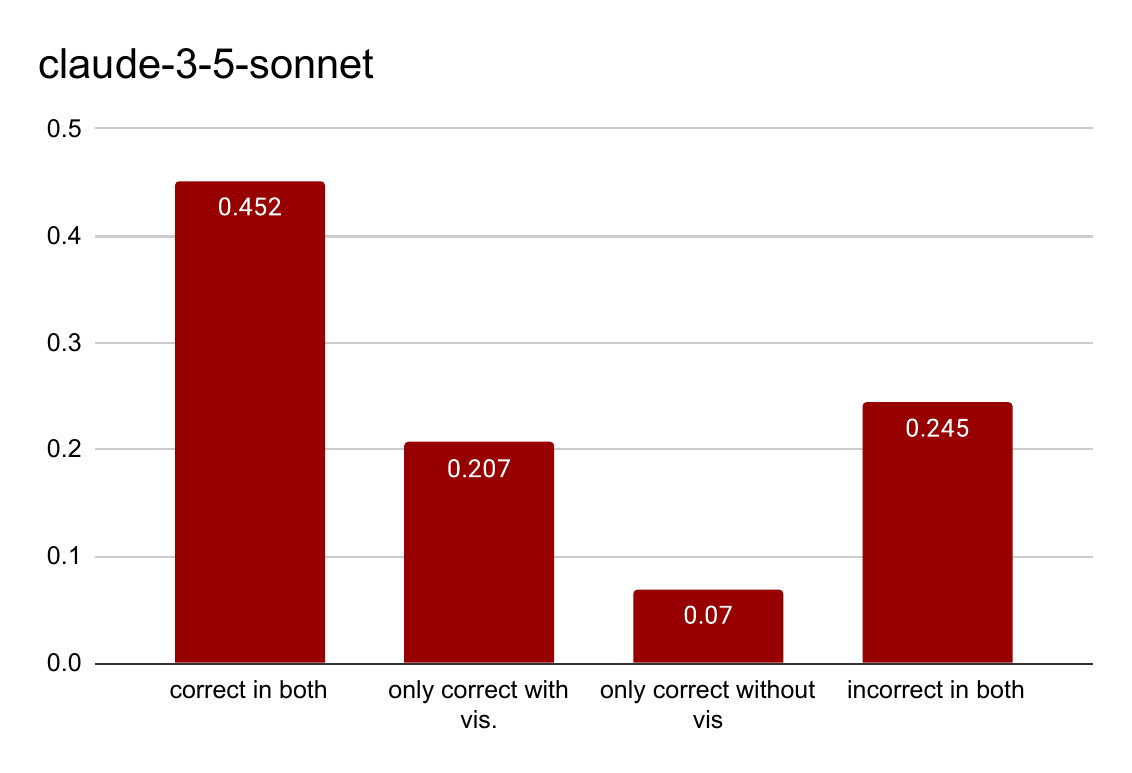}   
    \caption{The performance of the claude-3-5-sonnet mode during the question query with the VLAT dataset. Only 20\% of questions need visualization to answer them correctly.}
\label{fig:vlat_ablation_test_claude}
\end{figure}

We also experimented with claude-3-5-sonnet with and without visualization presented.
From Fig.~\ref{fig:vlat_ablation_test_claude}, we can tell that without the visualization presentation, the model can still answer many questions correctly without referring to the visualization.
Each experiment will create a new agent and each case can be tested five times.

\section{ChartQA Dataset}

The ChartQA dataset is curated from online sources.
The dataset mainly contains bar, line, and pie charts~\cite{hoque2022chart}. A standard example from the visualization is in Fig.~\ref{fig:chat_qa_data_example}. The related question is designed as:
``In the first quarter of 2021, what percentage of paid Microsoft advertising clicks originated from mobile devices?''.
During the evaluation, we use the following prompt: ``[Question] please answer the question in the format: Answer:"

\begin{figure}[!hpbt]
\centering   
    \includegraphics[width=\linewidth]{appendix/multi_col_535.png}   
    \caption{An example of the ChartQA dataset.}
\label{fig:chat_qa_data_example}
\end{figure}

\section{VILA Dataset}
\begin{figure*}[t]
\centering   
 \includegraphics[width=0.24\linewidth]{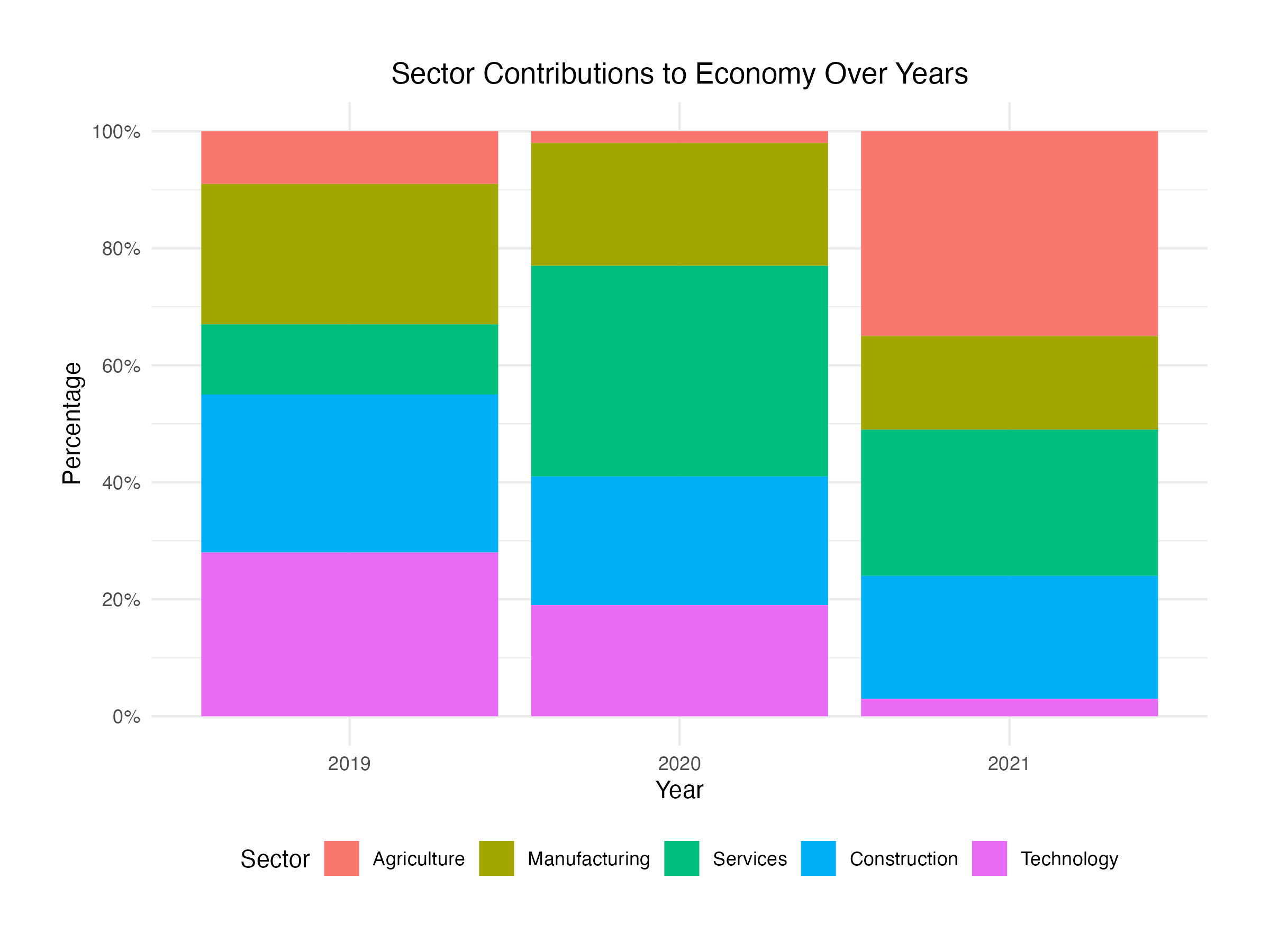}
 \includegraphics[width=0.24\linewidth]{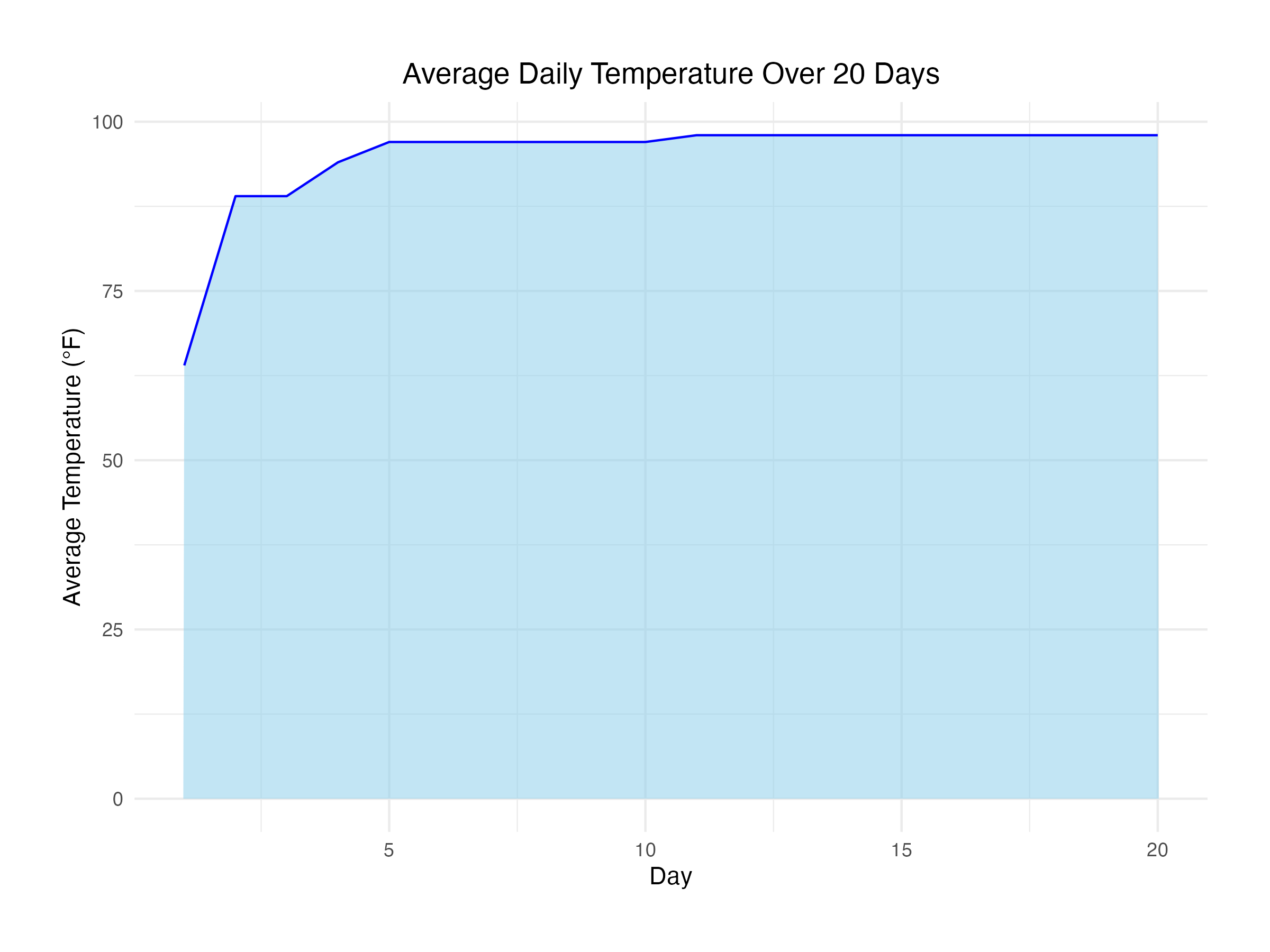}
 \includegraphics[width=0.24\linewidth]{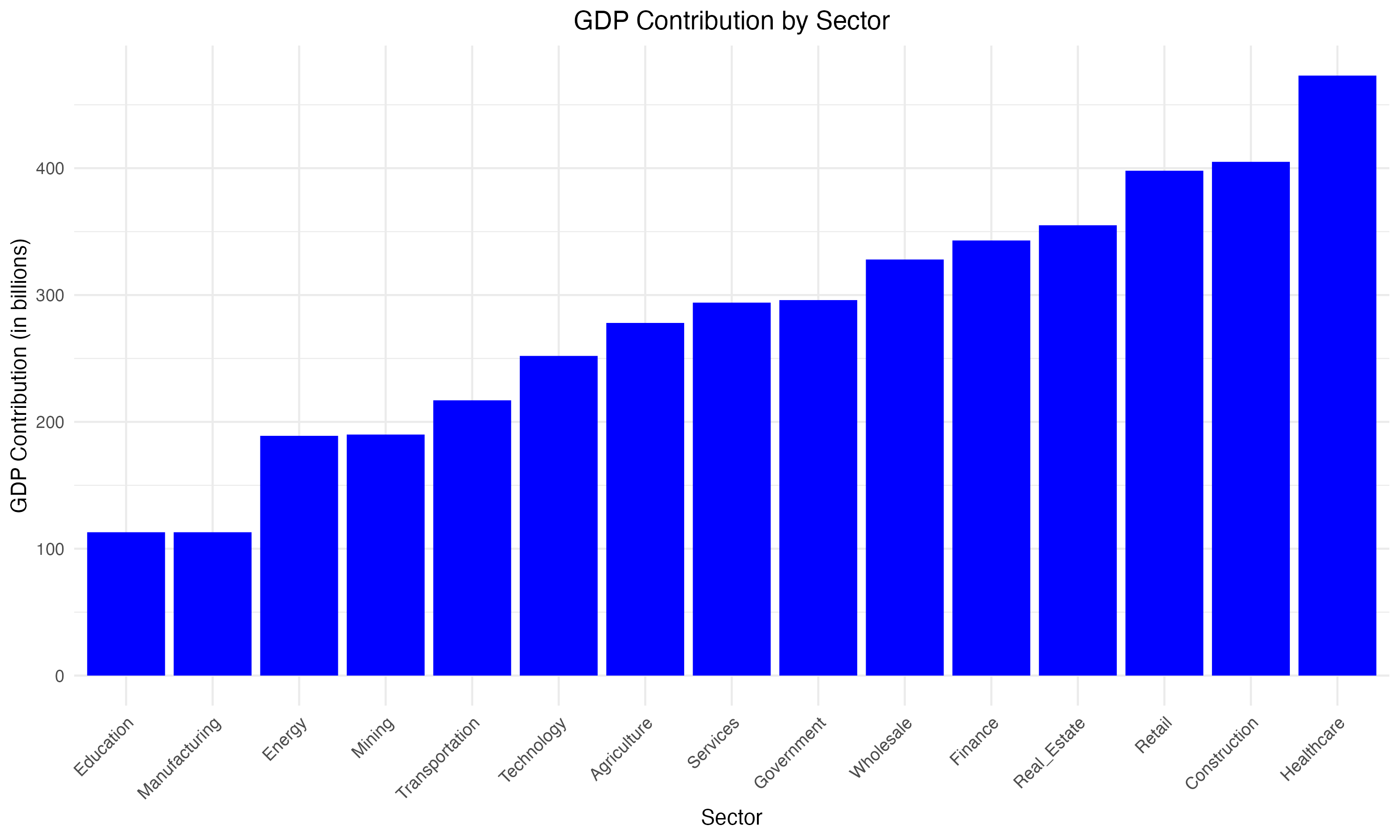}
 \includegraphics[width=0.24\linewidth]{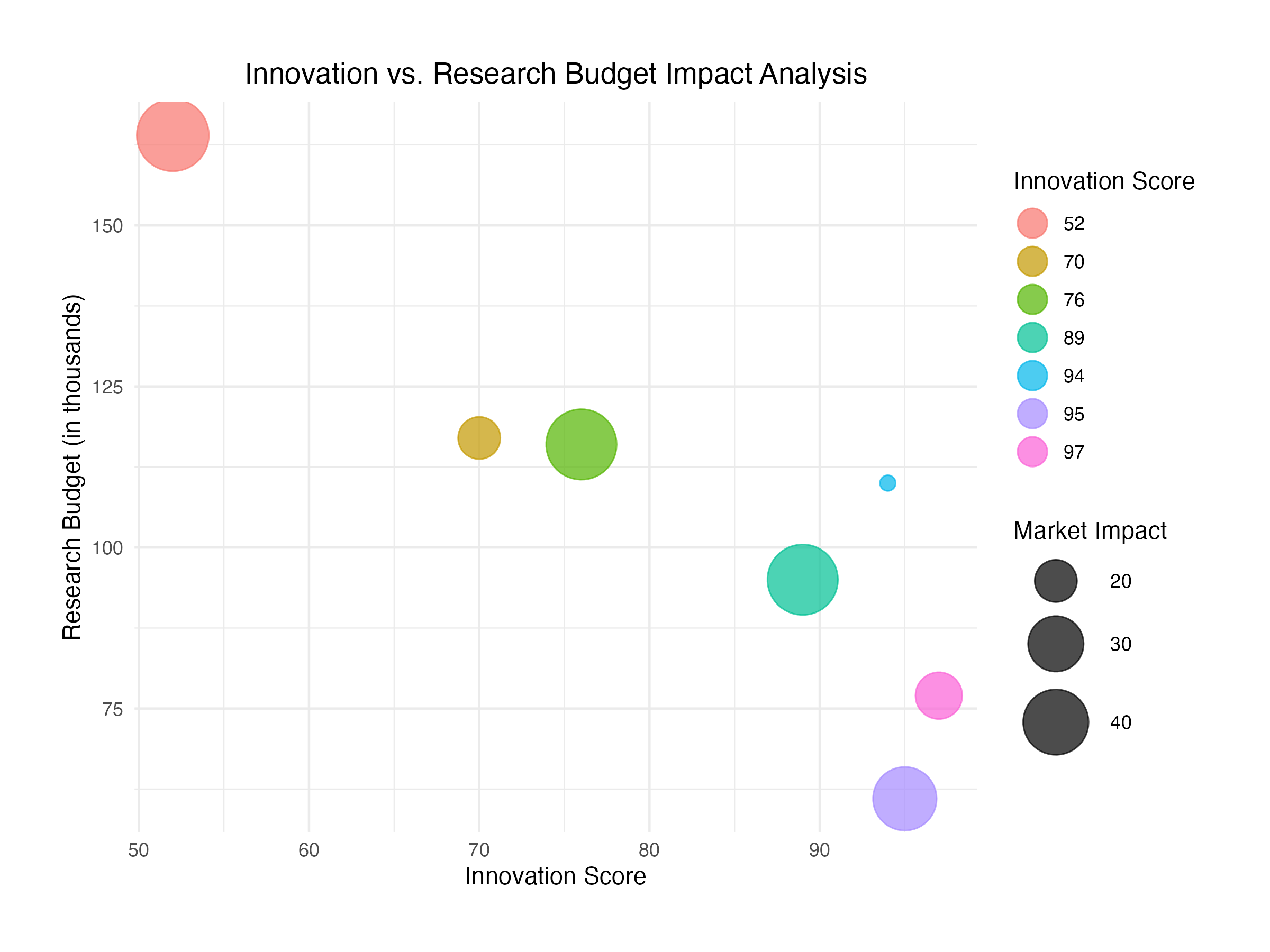}
 \includegraphics[width=0.24\linewidth]{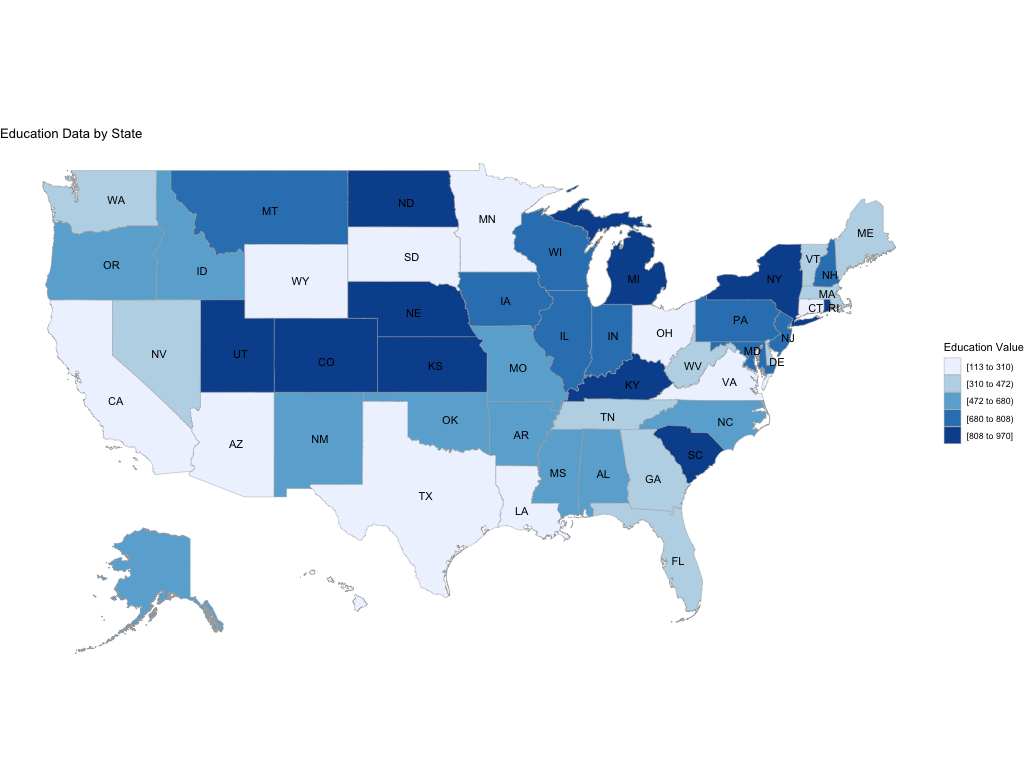}
 \includegraphics[width=0.24\linewidth]{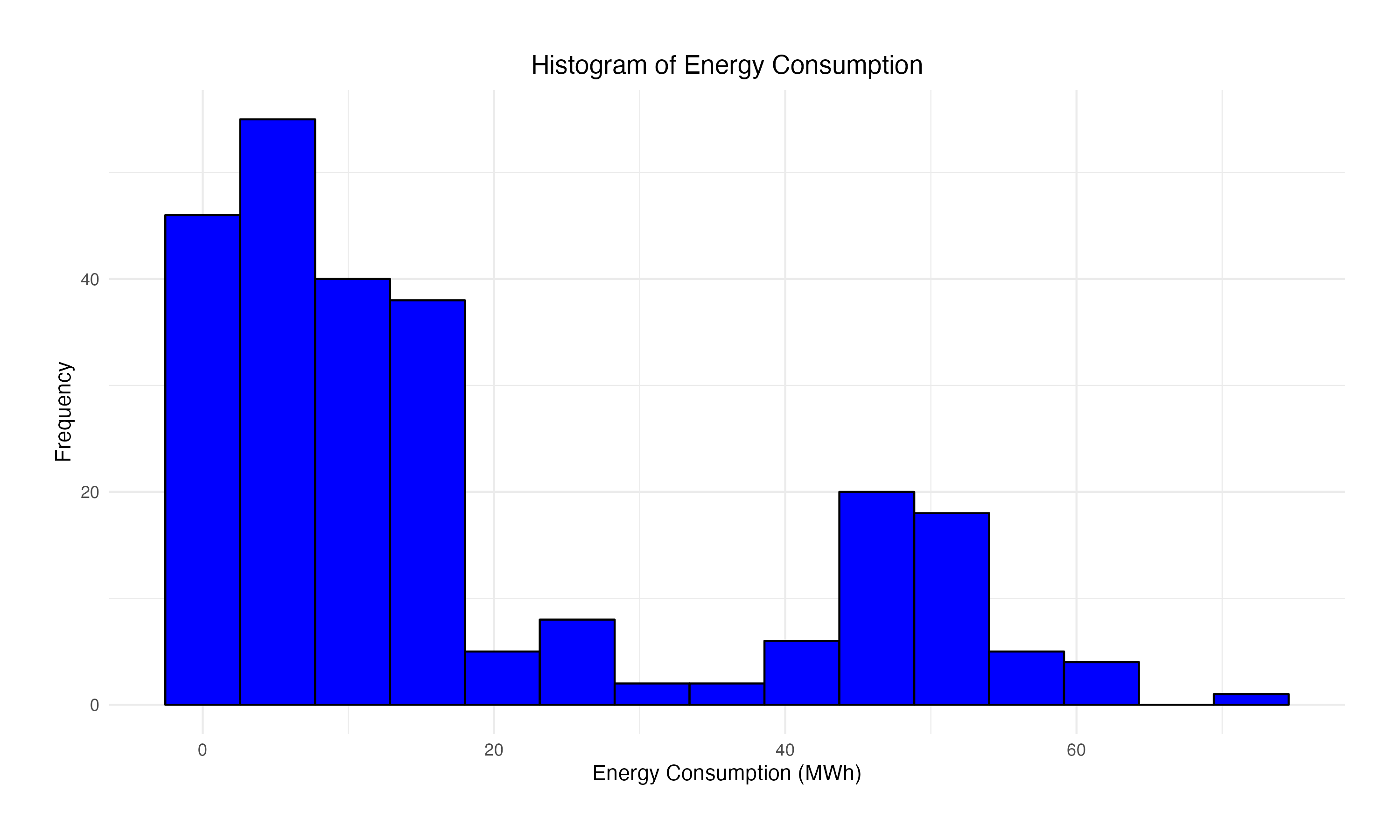}
 \includegraphics[width=0.24\linewidth]{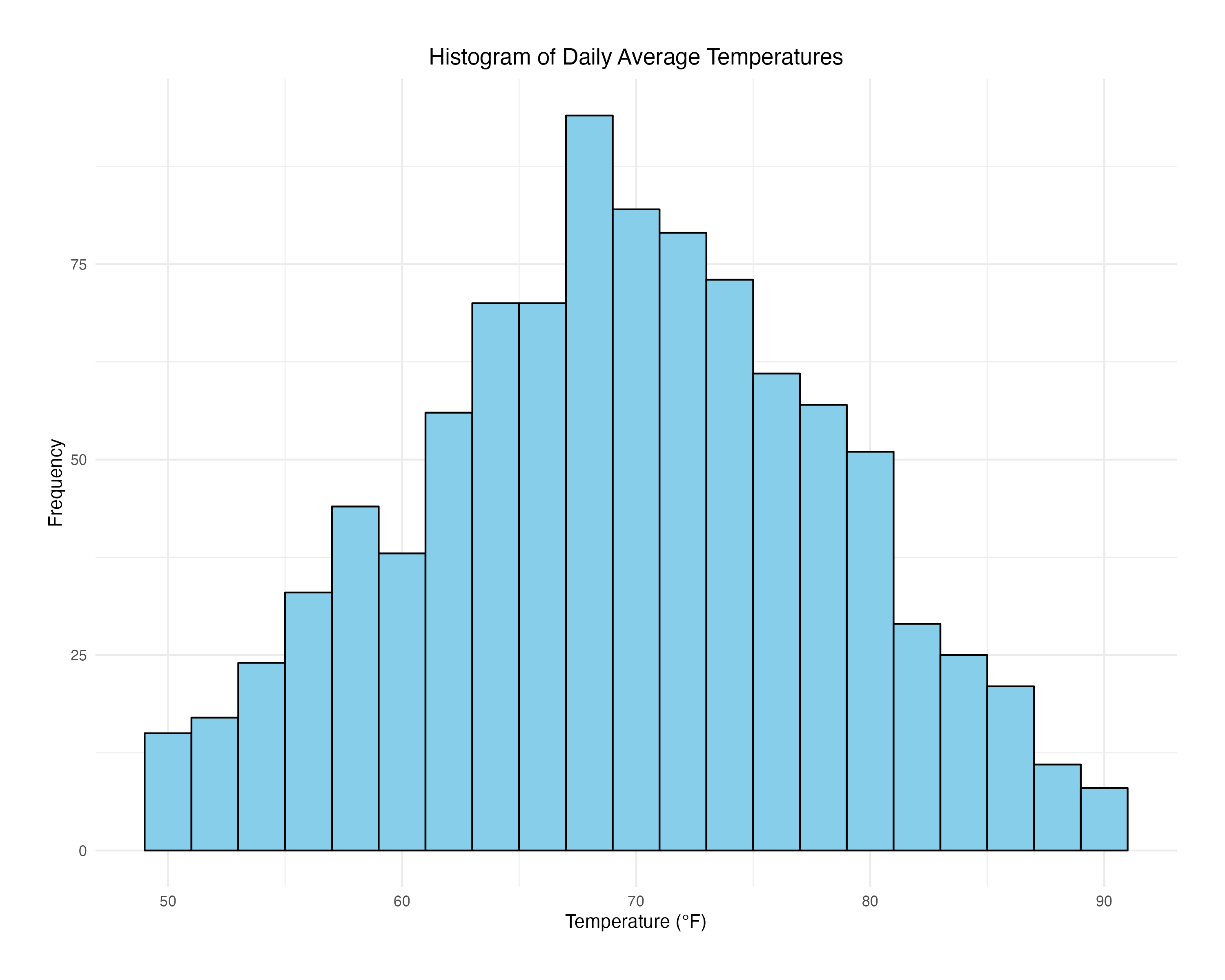}
 \includegraphics[width=0.24\linewidth]{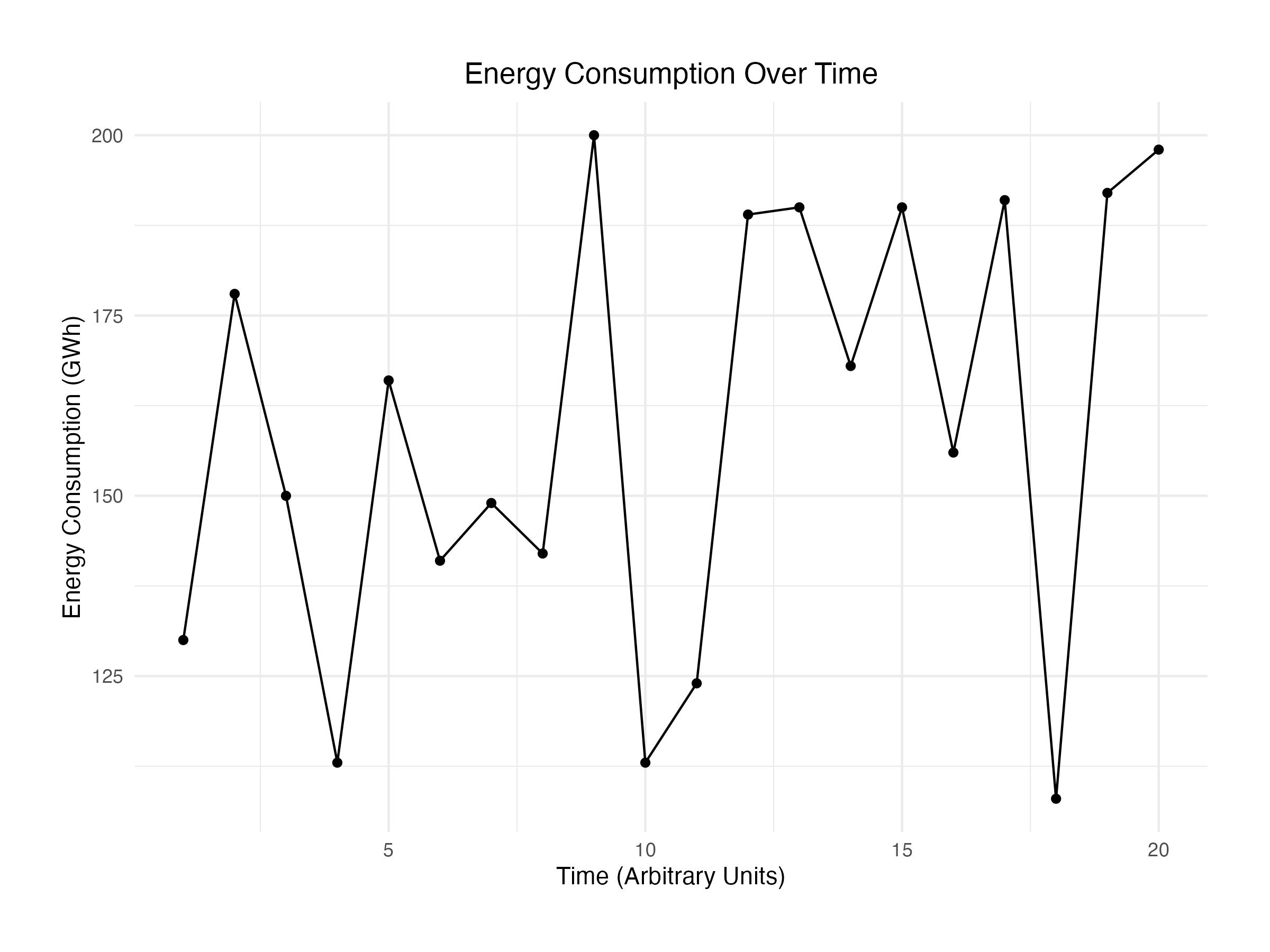}
 \includegraphics[width=0.24\linewidth]{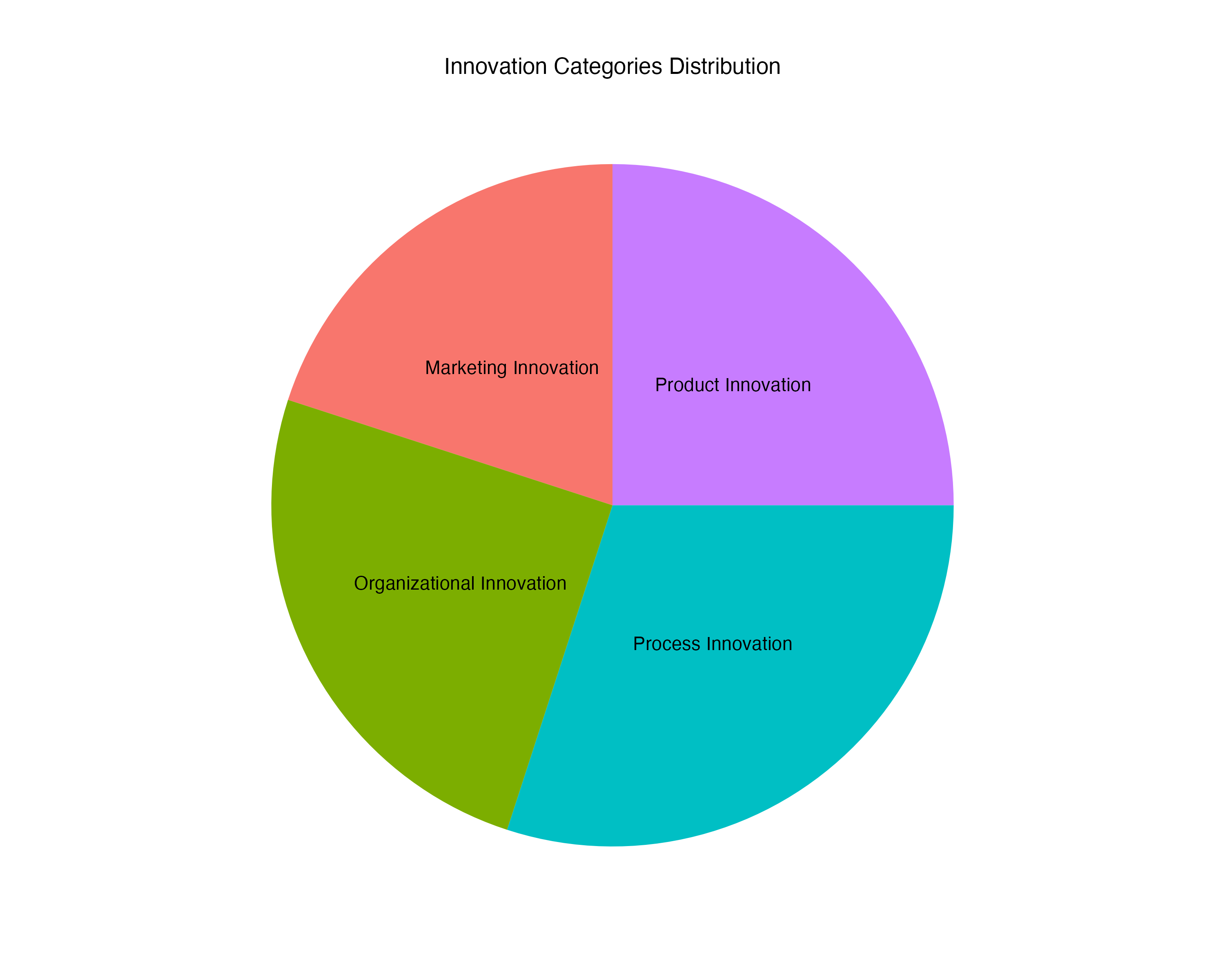}
 \includegraphics[width=0.24\linewidth]{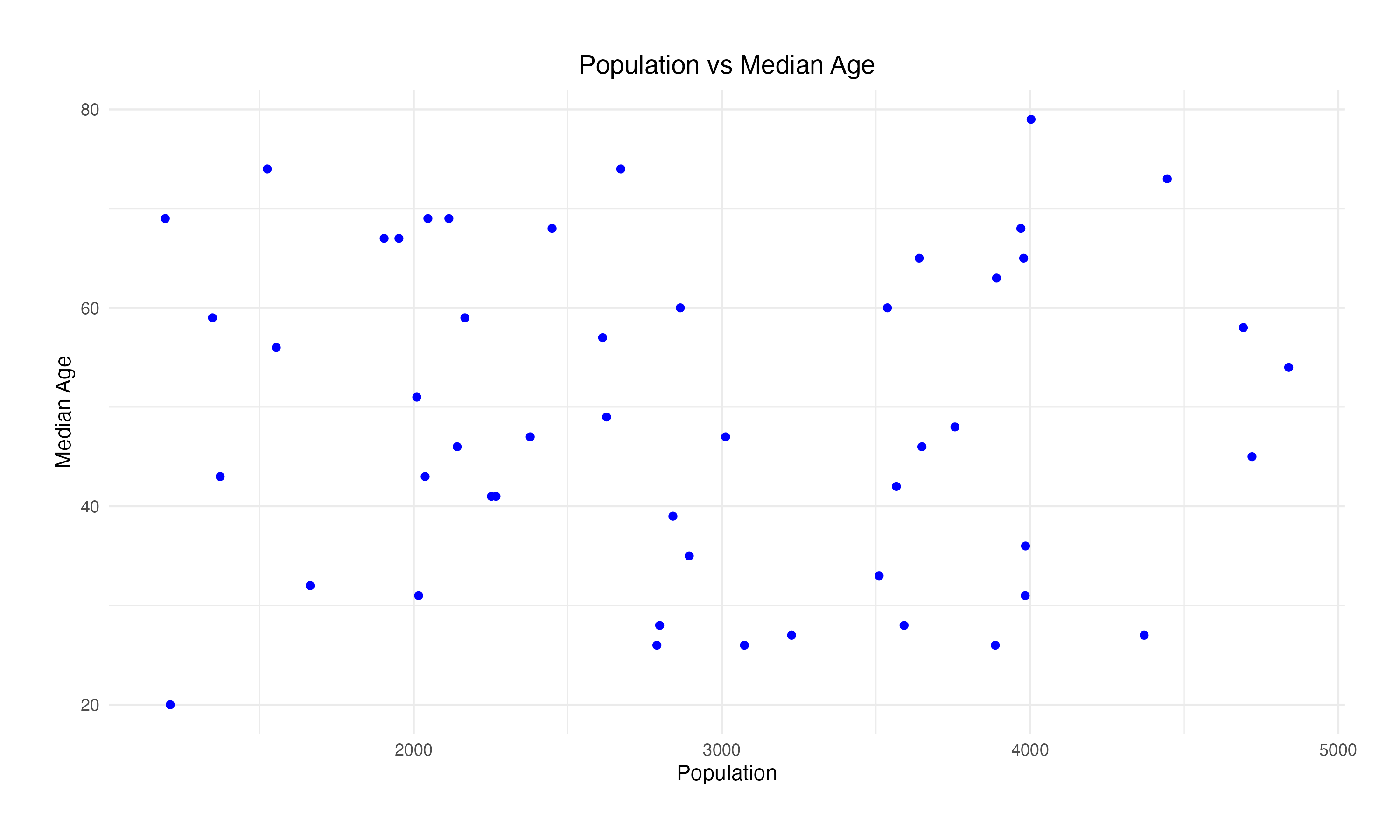}
 \includegraphics[width=0.24\linewidth]{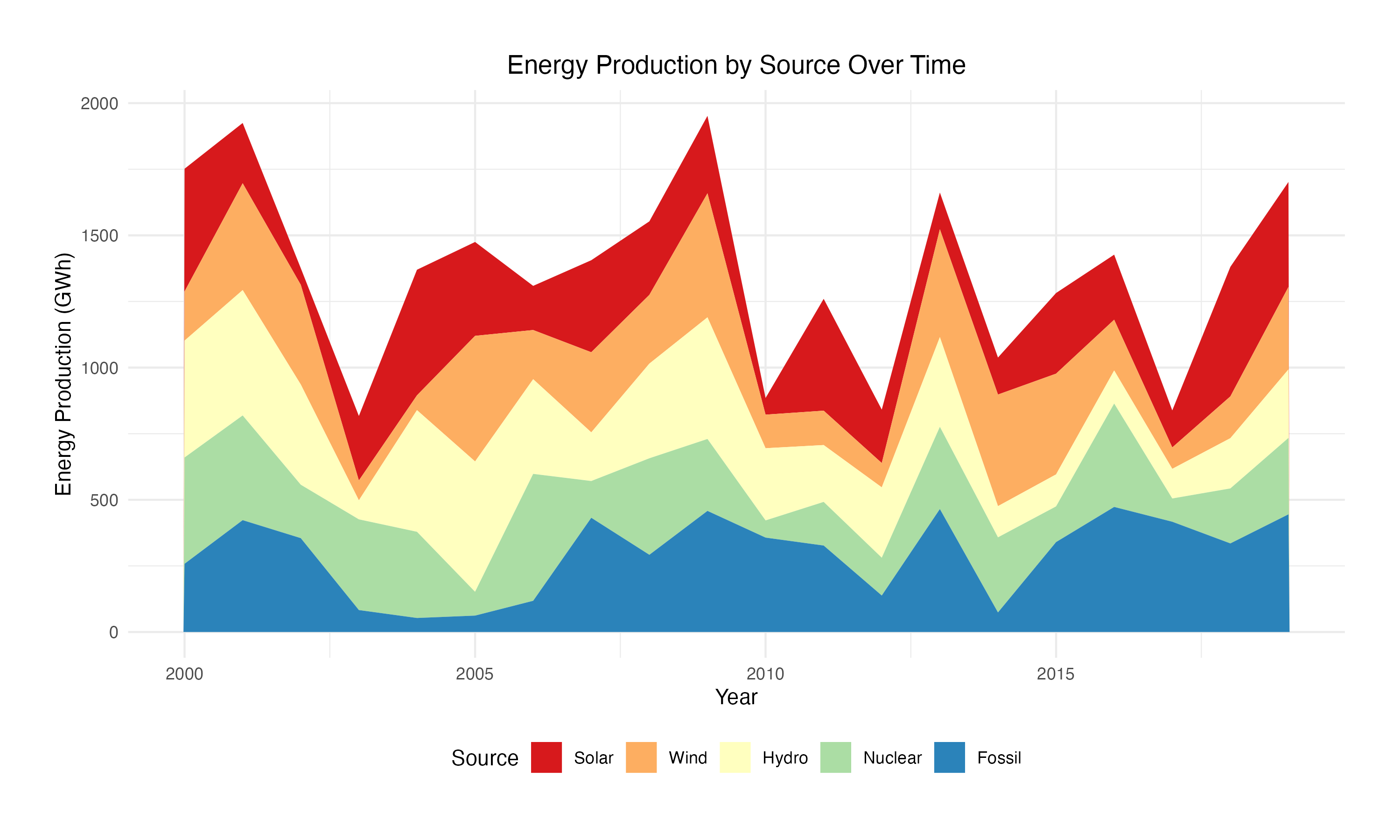}
 \includegraphics[width=0.24\linewidth]{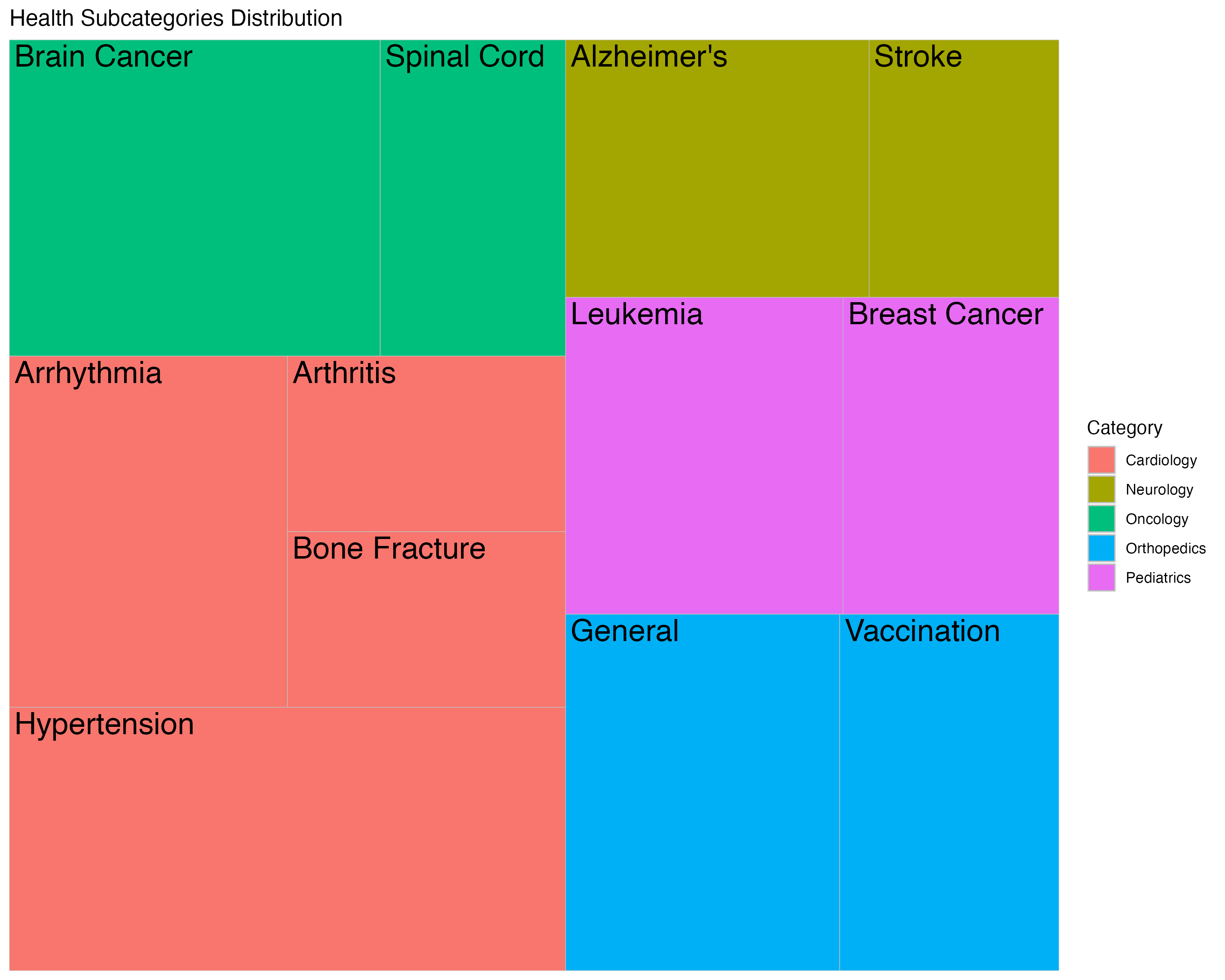}
 \caption{We use our proposed framework to perform the sanity check over the Vila dataset. Here demonstrates a small subset of the data.}
\label{fig:vila_dataset}
\end{figure*}

VILA~\cite{cui2024promises} Dataset contains 108 visualizations and 1108 questions. 
The dataset used the 12 visualizations in the VLAT dataset as a template visualization but with different configurations such as colors, axis configuration, and numbers of data items.
The visualization is generated through the R code produced by a large language model. 
The selected samples for the sanity check in this dataset can be referred to in the attached folder VILA\_Samples.
The dataset contains 13 different tasks with 8 different contexts. 
A subset of the detail visualization can be referred to in Fig.~\ref{fig:vila_dataset}.
The dataset is verified by two domain experts and tested over a large amount of participants.
We use the following prompt to query the language model:``[Question]  please pick one answer in the format: why? Answer: (A) or (B) or (C) or (D)".

\section{Sanity Check Table Over VLAT}
\begin{figure*}[t]
\centering   
 \includegraphics[width=0.24\linewidth]{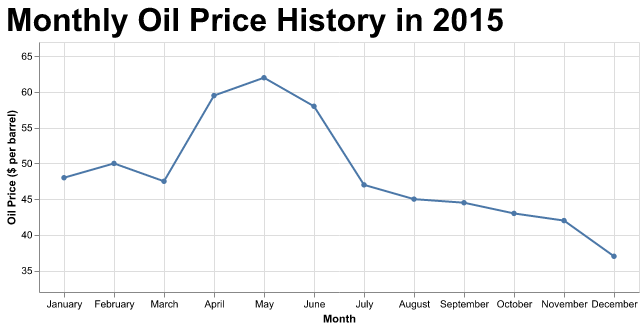}
 \includegraphics[width=0.24\linewidth]{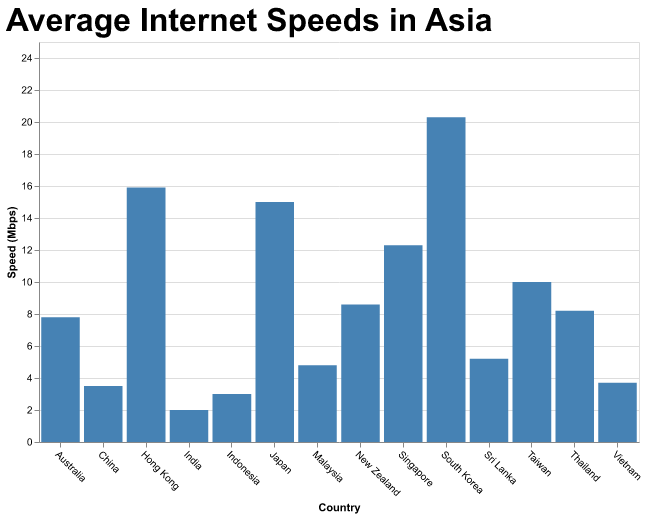}
 \includegraphics[width=0.24\linewidth]{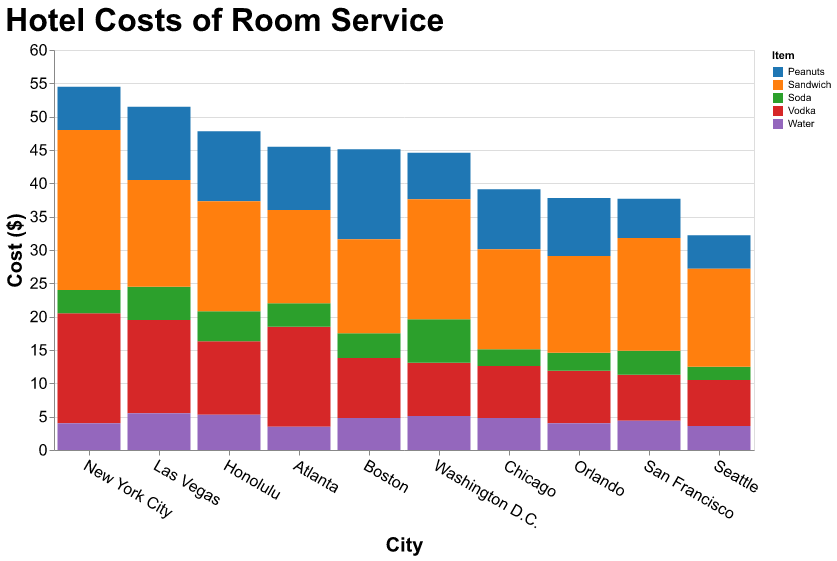}
 \includegraphics[width=0.24\linewidth]{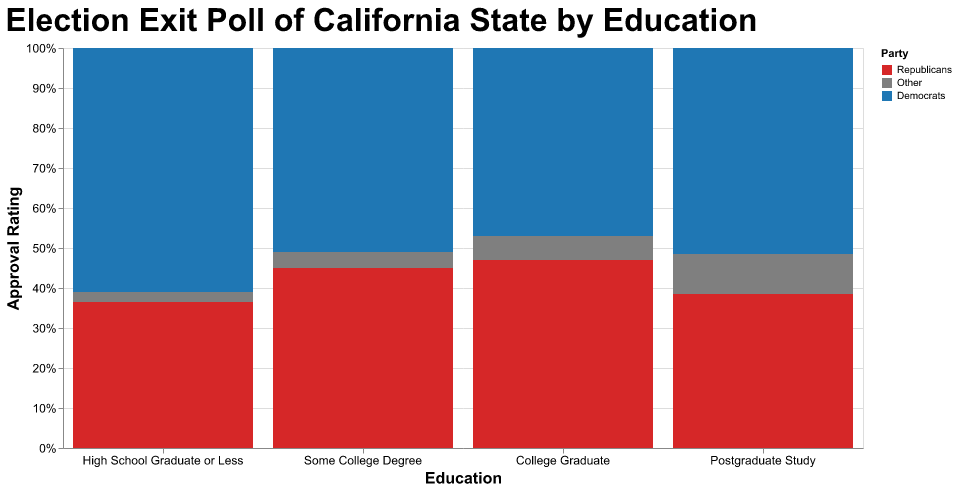}
 \includegraphics[width=0.24\linewidth]{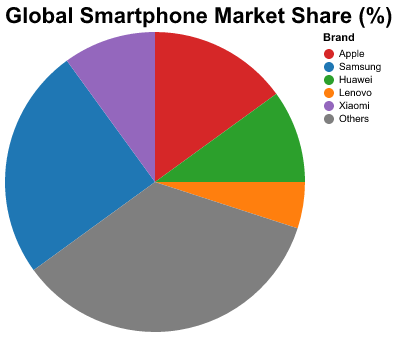}
 \includegraphics[width=0.24\linewidth]{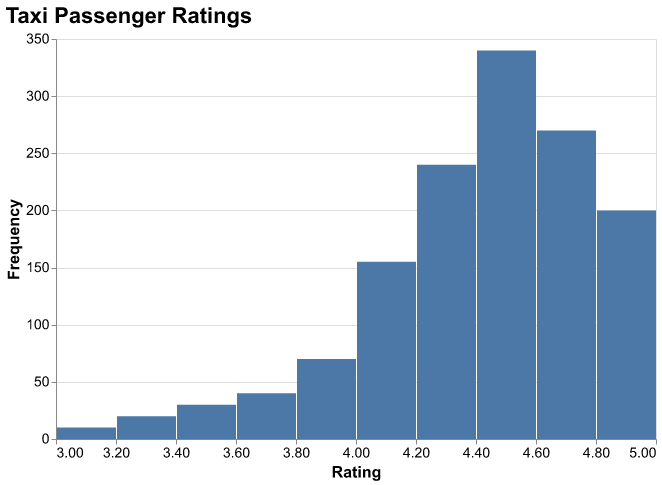}
 \includegraphics[width=0.24\linewidth]{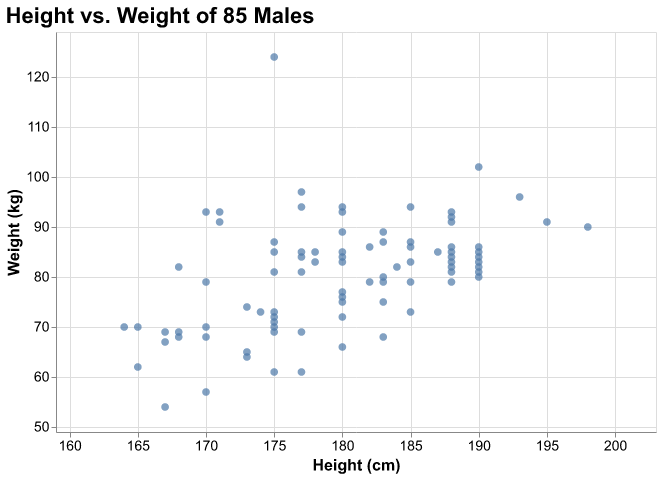}
 \includegraphics[width=0.24\linewidth]{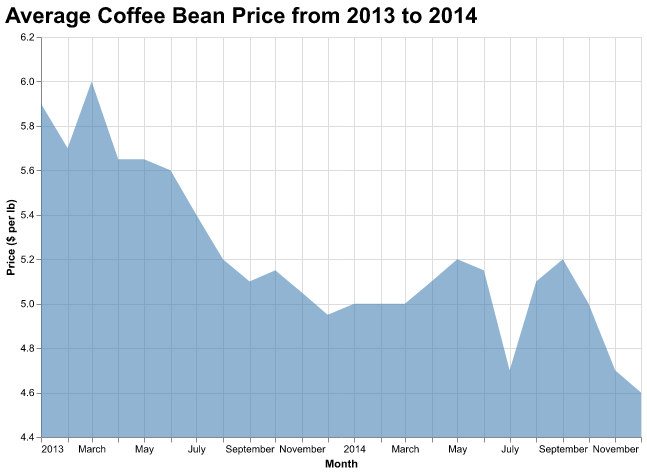}
 \includegraphics[width=0.24\linewidth]{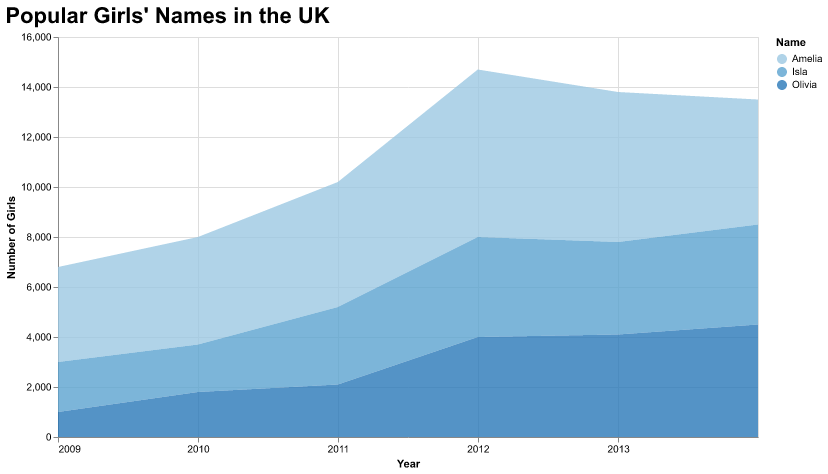}
 \includegraphics[width=0.24\linewidth]{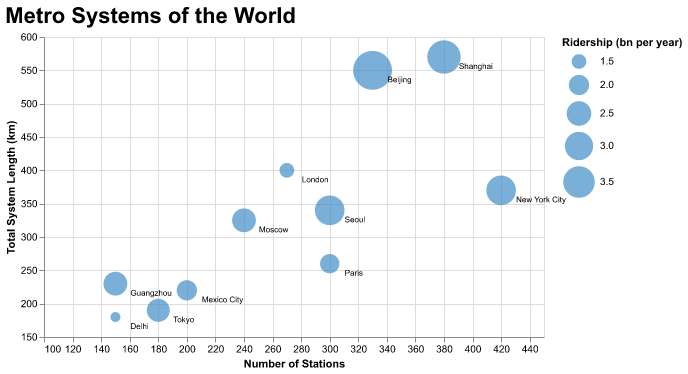}
 \includegraphics[width=0.24\linewidth]{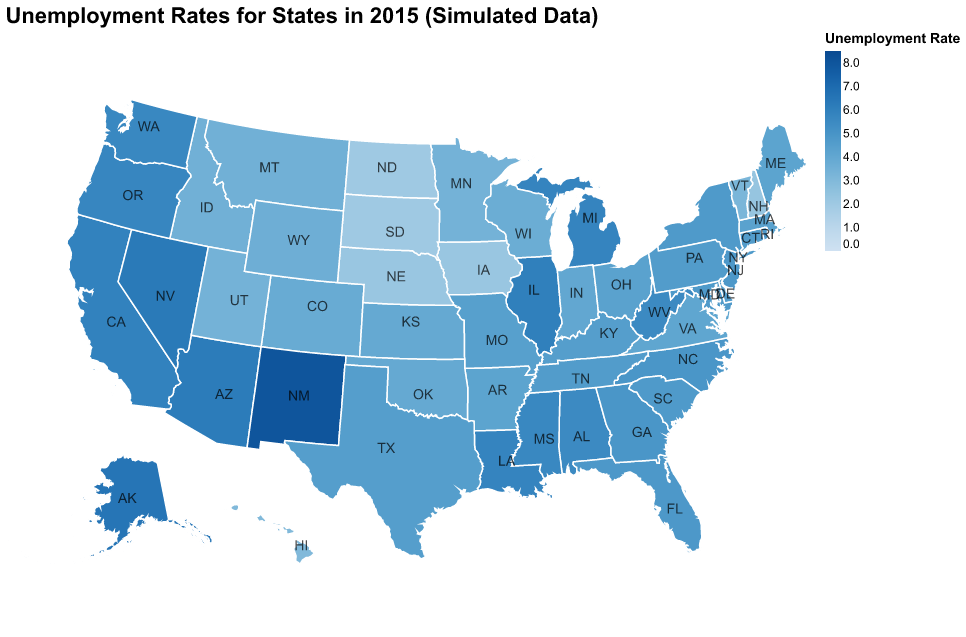}
 \includegraphics[width=0.24\linewidth]{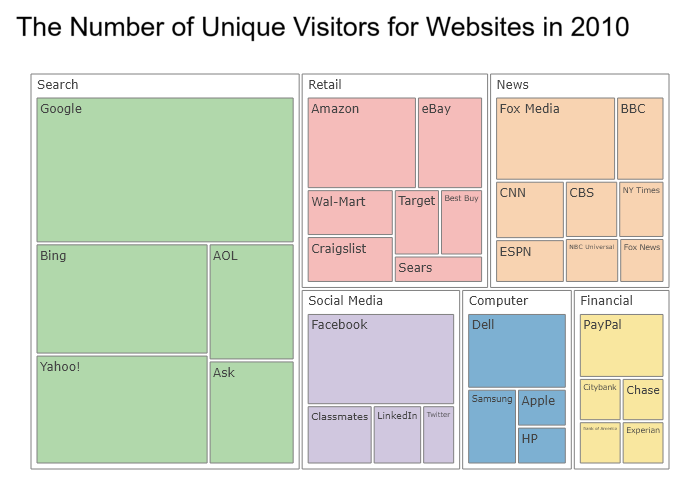}
 \caption{We use our proposed framework to perform the sanity check over the VLAT dataset.}
\label{fig:vlat_dataset}
\end{figure*}

Twelve visualizations used for evaluation can be referred to Fig.~\ref{fig:vlat_dataset} and the full sanity check table can refer to  Table~\ref{tab:vlat_question_data_table_sanity_check}.

\begin{table*}[t]
\setlength{\tabcolsep}{2.5pt} 
\resizebox{\textwidth}{!}{
\scriptsize
\centering
\begin{tabular}{ccccccccccl}
 \hline
Visualization  &   Task & P1 (See + Recall) & P2 (See + No Recall) & P3 (No See + Recall) & P4 (No Recall + No See) & C1 & C2 & C3\\  
\hline
{}                 & retrieve value & 0.2& 0.1& 0.03& 0.2&- &-&-\\
{}                 & find extremum & 1.0& 0.96& 0& 0.03&- & - & -\\
Line Chart (1)     & determine range & 0.666& 0.466& 0.966& 0.9 &  \checkmark & -  & - \\
{}                 & find trend & 1.0& 1.0& 0.666& 0.33&- &  \checkmark & -\\
{}                 & make comparison & 0.766& 0.533& 0.1& 0.233&-&-&-\\
\hline
{}                 & retrieve value & 0.6& 0.233& 0.2& 0.133& - & \checkmark & -\\
Bar Chart(2)   & find extremum & 1.0& 1.0& 1.0& 0.533&-& \checkmark&-\\
{}                 & determine range & 0.133& 0.2& 0.166& 0.06&-&-&-\\
{}                 & make comparison & 0.533& 0.433& 0.33& 0.3&-&-&-\\
\hline
{}                 & retrieve value & 0.96&        0.233&                0.2& 0& - & \checkmark & - \\
{}                 & retrieve value (rel.) & 0.56&        0.06&                0.16& 0.36& - & \checkmark & -\\
Stacked Bar Chart(3)  & find extremum & 0.03&        0.1&                0& 0&-&-&-\\
{}                 & make comparison & 0.06&        0&                0& 0&-&-&-\\
{}                 & make comparison (rel.) & 0.366&        0.466&                0.766& 0.9& \checkmark &  - & - \\
\hline
{}                 & retrieve value (rel.) & 1.0&        0.16&                1.0& 0.1& - & \checkmark &  -\\
100\% Stacked Bar Chart(4)  & find extremum (rel.) & 0.0&        0.06&                0.0& 0.4&-&-&-\\
{}                 & make comparison (rel.) & 0.533&        0.6&                0.93& 0.6&-& \checkmark&\\
\hline
{}                 & retrieve value (rel.) & 0.633&        0.766&                1.0& 0.2&-& \checkmark&-\\
Pie Chart(5)      & find extremum (rel.) & 1.0&        0.266&                1.0& 0.233& - & \checkmark & - \\
{}                 & make comparison (derived) & 0.93&        1.0&                0.93& 0.03&-& \checkmark &-\\
\hline
{}                 & retrieve value (derived) & 0.166&        0.33&                0.533& 0.466& \checkmark &  - & - \\
Histogram(6)   & find extremum (derived) & 0.833&        0.966&                0.0& 0.06&-&-&-\\
{}                 & make comparison (derived) & 0.833&        0.433&                0& 0&  \checkmark &-& - \\
\hline
{}                 & retrieve value & 0.0&        0.46&                0.0& 0.23& \checkmark &-& - \\
{}                 & find extremum & 0.03&        0&                0& 0&-&-&-\\
{}                 & determine range & 0.866&        0.633&                0.5& 0.33&\checkmark& - & -\\
Scatterplot(7)  & find anomalies & 0&        0.1&                0& 0&-&-& -\\
{}                 & find clusters & 0.733&        0.233&                0.033& 0.066& - & \checkmark & - \\
{}                 & find correlation & 1.0&        1.0&                1.0& 0.56& - &  \checkmark & - \\
{}                 & make comparison & 1.0&        1.0&                1.0& 0.93& \checkmark &  - & - \\
\hline
{}                 & retrieve value & 0.233&        0.433&                0.266& 0.433& - & - & -\\
Area Chart(8)  & find extremum & 0.03&        0.93&                0& 0& - & - & \checkmark \\
{}                 & determine range & 0.5&        0.56&                0.3& 0.36&-&- & -\\
{}                 & find trend & 1.0&        0.93&                0.66& 0.13&- &  \checkmark & -\\
\hline
{}                 & retrieve value & 0.7&        0.1&                0.53& 0.3& - &  \checkmark &  -\\
{}                & retrieve value (rel.) & 0.6&        0&                0.3& 0&  - & \checkmark &  - \\
Stacked Area Chart(9)  & find extremum & 1.0&        0.96&                0.5& 0.26&- &  \checkmark & -\\
{}                 & find trend & 0.2&        0.23&                0.06& 0.06&-&-&-&\\
{}                 & make comparison & 1.0&        1.0&                0.96&0.8& \checkmark &  - & -\\
{}                 & make comparison (rel.) & 1.0&        0.96&                0.96& 0.9& \checkmark &  - & -\\
\hline
{}                 & retrieve value & 0.166&        0.833&                0.03& 0.53& \checkmark &- &  -\\
{}                 & find extremum & 0.4&        0.96&                1.0& 0.7&  \checkmark &  - & -\\
{}                 & determine range & 0&        0.03&                0.0& 0.2&- & -& -\\
Bubble Chart(10)        & find anomalies & 0.76&        0.6&                0.63&0.16& - &  \checkmark & -\\
{}                 & find clusters & 0.93&        0.93&                0.4& 1.0& \checkmark & - & -\\
{}                 & find trend & 0.06&        0.43&                0& 1.0&  \checkmark & - &  -\\
{}                 & make comparison & 0.66&        0.83&                0.06& 0.96& \checkmark & - & - \\
\hline
{}                 & retrieve value & 0.1&        0.53&                0.0& 0.76& \checkmark & - &  -\\
Choropleth Map(11)          & find extremum & 1.0&        1.0&                1.0&0.8&  \checkmark &  - &  -\\
{}                 & make comparison & 0.66&        0.56&                0.3& 0.86&  \checkmark & - & -\\
\hline
{}                 & find extremum (rel.) & 1.0&        0.43&                1.0& 0.3&  - &  \checkmark &  -\\
Treemap(12)         & make comparison (rel.) & 1.0&        1.0&                0.26& 0.0& - & - &  -\\
{}                 & identify hierarchical structure & 1.0&        1.0&                1.0& 1.0&  \checkmark &  - & -\\
 \hline

\end{tabular}
}
\caption{The table displays the sanity check of MLLM over each question of the VLAT datasets. A rule is broken only if the observation is statistically significant.}
\label{tab:vlat_question_data_table_sanity_check}
\normalsize
\end{table*}

\section{VLATforge and Counterfactual Dataset}

\begin{figure*}[t]
\centering   
 \includegraphics[width=0.24\linewidth]{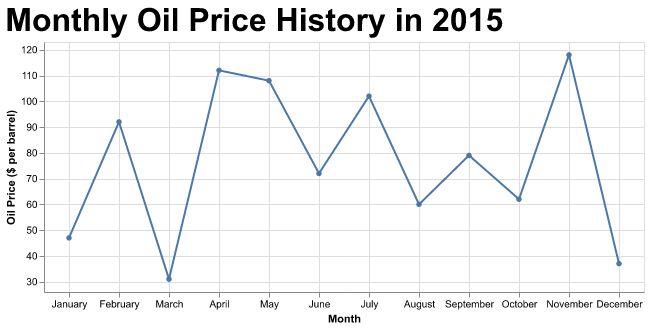}
 \includegraphics[width=0.24\linewidth]{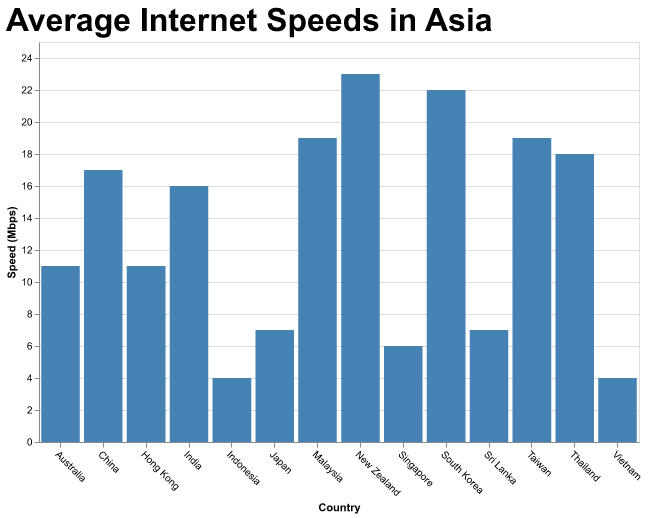}
 \includegraphics[width=0.24\linewidth]{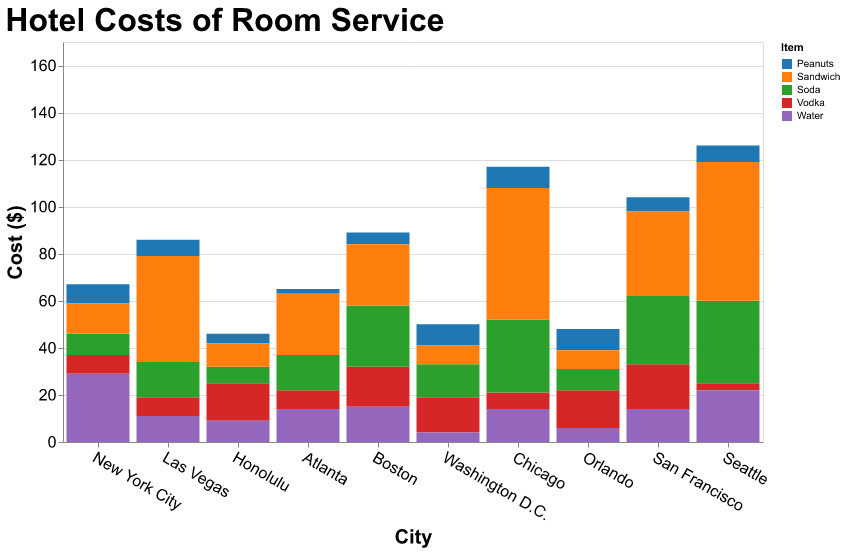}
 \includegraphics[width=0.24\linewidth]{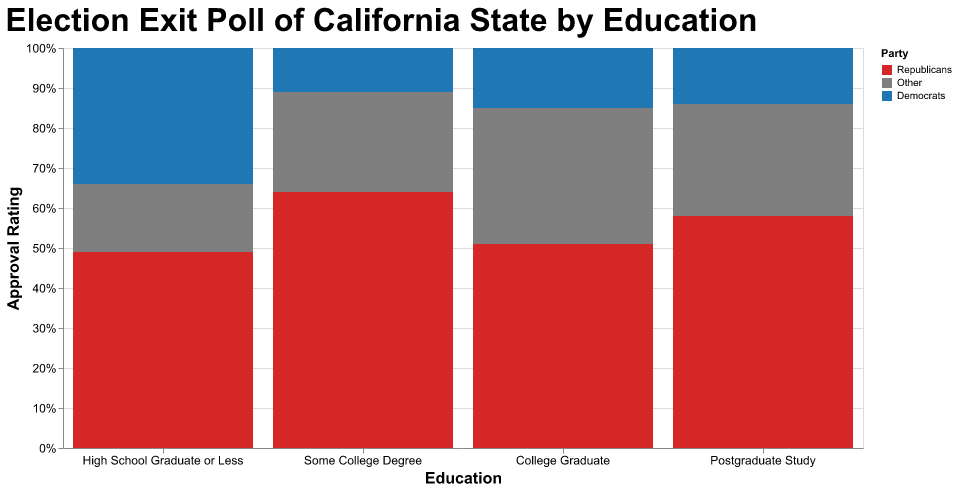}
 \includegraphics[width=0.24\linewidth]{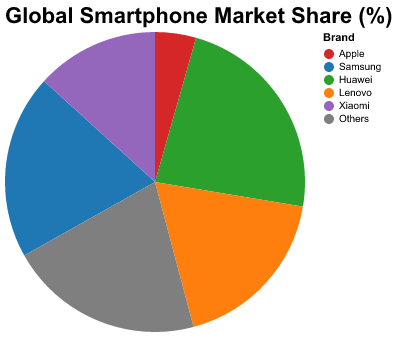}
 \includegraphics[width=0.24\linewidth]{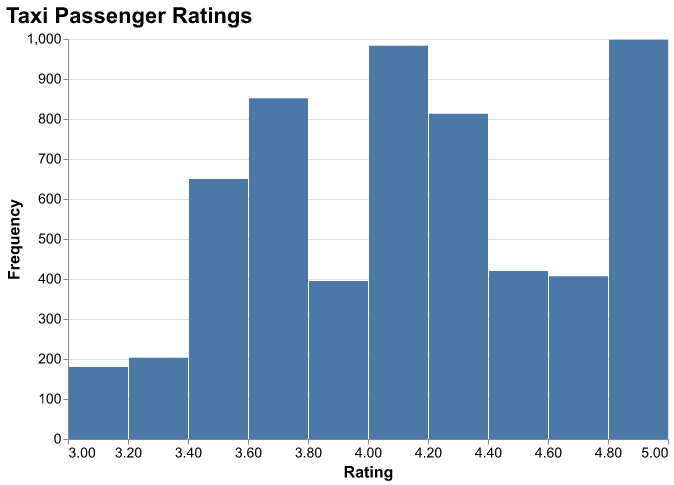}
 \includegraphics[width=0.24\linewidth]{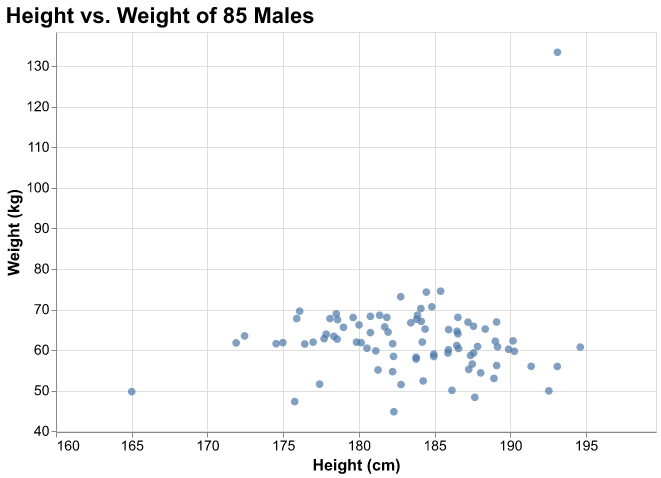}
 \includegraphics[width=0.24\linewidth]{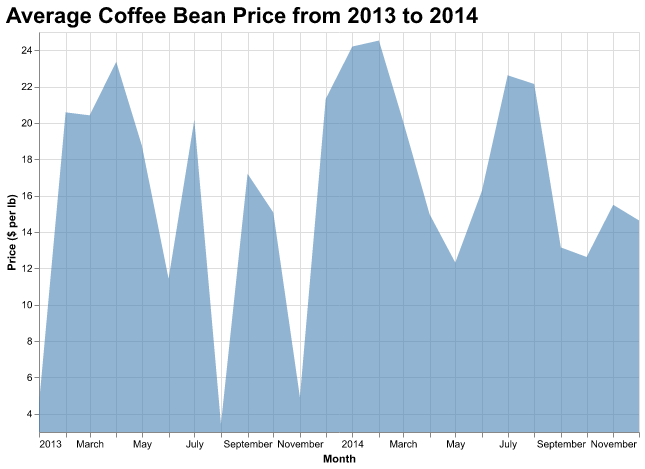}
 \includegraphics[width=0.24\linewidth]{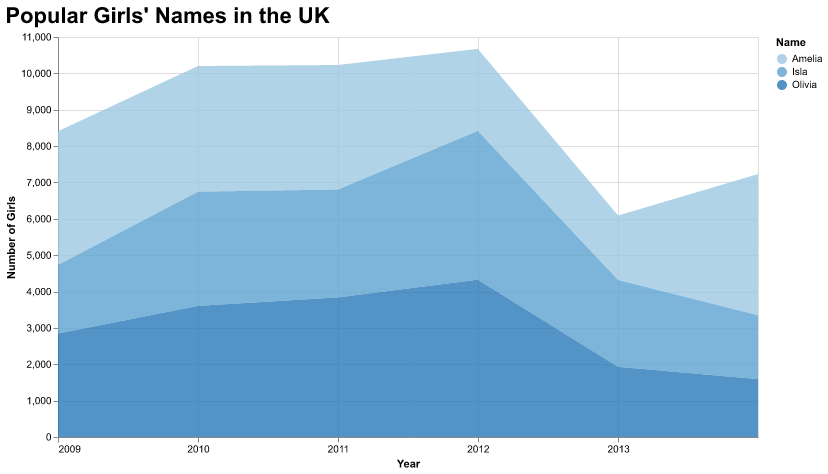}
 \includegraphics[width=0.24\linewidth]{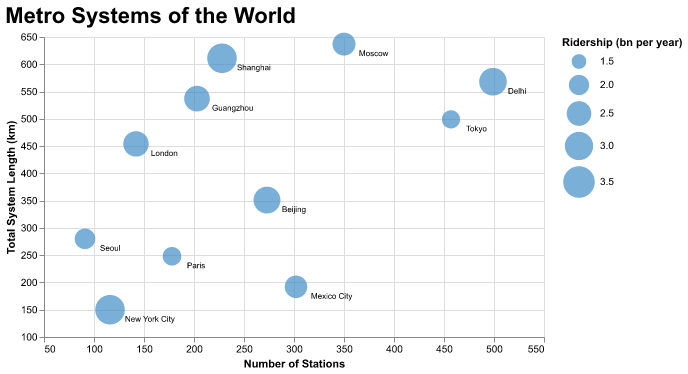}
 \includegraphics[width=0.24\linewidth]{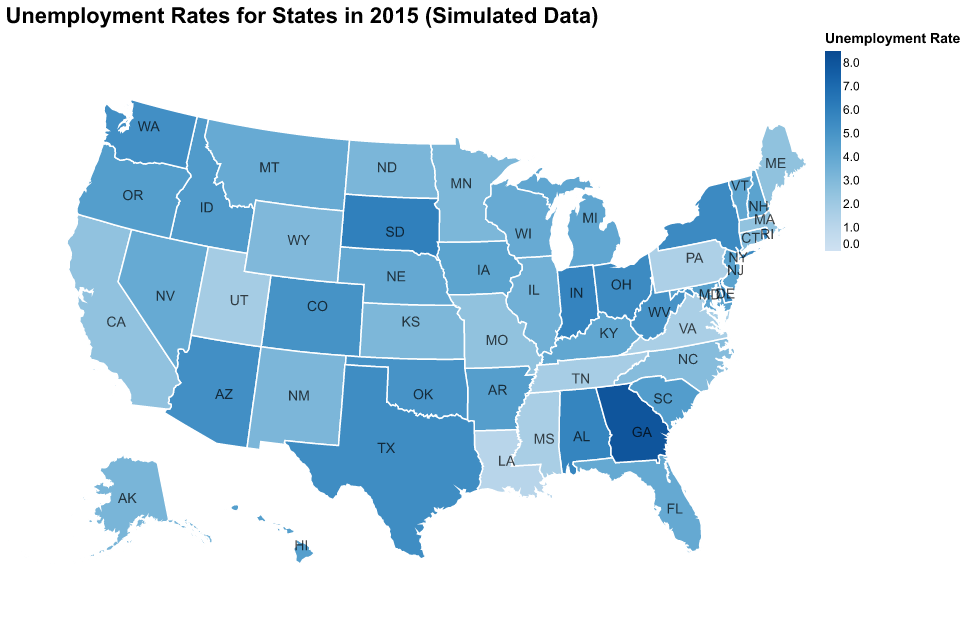}
 \includegraphics[width=0.24\linewidth]{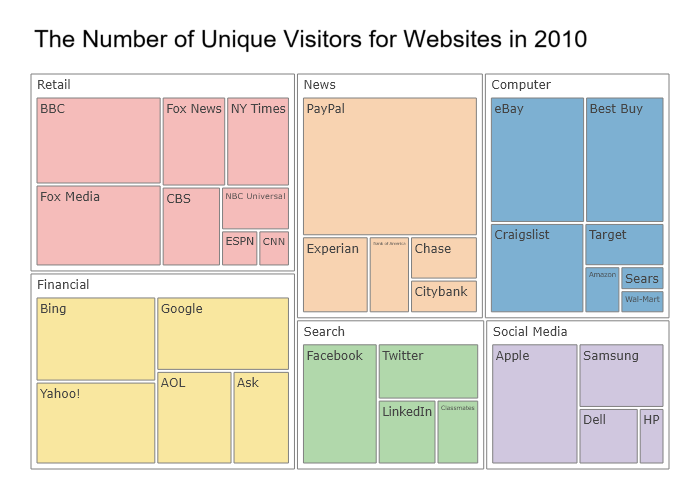}
 \caption{We use our proposed framework to perform the sanity check over the VLATforge dataset.}
\label{fig:vlatforge_dataset}
\end{figure*}

VLATforge uses the context information from VLAT. 
However, the data used to construct the visualization is randomly generated.
Therefore, the visualization information does not match with the real-world information.
The information presented in the visualization is counterfactual.
The visualization generated by the VLATforge can be referred to Fig.~\ref{fig:vlatforge_dataset}.
The following describes the data range when we generate the VLATforge benchmark.

\textbf{Question 1:}
The data on the oil price is randomly generated between 30 to 120.

\textbf{Question 2:}
The data of the Internet speed is generated between 0 to 25.

\textbf{Question 3:}
The data  for each room server is randomly generated with the following range:
vodka: 2 - 30, soda: 2 - 40, peanuts: 2 - 10, water: 2 - 30, sandwich: 2 - 60.

\textbf{Question 4:}
democrats: 10\% - 50\%, other: 10\% - 40\%, republic: 1 - democrats - other.

\textbf{Question 5:}
market share for each category: 2 - 60 

\textbf{Question 6:}
Frequency: 5 - 1000

\textbf{Question 7:}
We create a distribution that is similar to the original VLAT. For data, each point will be randomly scaled up in weight and height with 0\%-10\%

\textbf{Question 8:}
The coffee price is randomly generated with a range 3 - 25

\textbf{Question 9:}
the number of populations in each category. amelia: 1500 - 4500, Isla: 1500 - 4500, Bolivia: 1500 - 4500

\textbf{Question 10:} the number of parameters
km: 150 - 650, station: 80 - 500, ridership per year: 1.5, 3.5

\textbf{Question 11:}
unemployment: 1.0 - 6.0 and 
the maximum is 8.0

\textbf{Question 12:}
The number of visitors for each category is 50 - 999.
During the generation, the maximum value is: 2000
We also randomly shift the category that the company belongs to.

\end{appendices}

\end{document}